\documentclass[notitlepage,prd,showkeys,preprintnumbers,floatfix,superscriptaddress,nofootinbib]{revtex4-1}
\usepackage{float}
\usepackage{amsmath,amssymb}
\usepackage{graphicx}
\usepackage{mathtools,slashed}
\usepackage{bm}
\usepackage{comment}
\usepackage{xcolor}
\usepackage{subfigure}
\usepackage{array}
\usepackage{multirow}
\usepackage{url}
\usepackage{color}
\usepackage[T1]{fontenc}
\usepackage[latin9]{inputenc}
\usepackage{graphicx}
\usepackage{esint}
\usepackage{hyperref}
\usepackage{atbegshi}
\usepackage{lipsum}
\usepackage{booktabs}

\begin{document}

\preprint{\vbox{\hbox{DESY 21-201, HU-EP-21/49}} }

\title{Inclusive rates from smeared spectral densities\\ in the two-dimensional O(3) non-linear $\sigma$-model}

\author{John Bulava}
\email[]{john.bulava@desy.de}
\affiliation{
Deutsches Elektronen-Synchrotron DESY,
Platanenallee 6, 15738 Zeuthen, Germany
}

\author{Maxwell T. Hansen}
\email[]{maxwell.hansen@ed.ac.uk}
\affiliation{
Higgs Centre for Theoretical Physics, School of Physics and Astronomy,
The University of Edinburgh, Edinburgh EH9 3FD, UK
}

\author{Michael W. Hansen}
\email[]{michael.hansen@uni-graz.at}
\affiliation{University of Graz, Institute for Physics, A-8010 Graz, Austria
}

\author{Agostino Patella}
\email[]{agostino.patella@physik.hu-berlin.de}
\affiliation{
Institut f\"{u}r Physik und IRIS Adlershof, Humboldt-Universit\"{a}t zu Berlin,
Zum Gro{\ss}en Windkanal 6, D-12489 Berlin, Germany
}

\author{Nazario Tantalo}
\email[]{nazario.tantalo@roma2.infn.it}
\affiliation{
University and INFN of Roma Tor Vergata, Via della Ricerca Scientifica 1, I-00133, Rome, Italy
}

\date{\today}

\begin{abstract}
	This work employs the spectral reconstruction approach of Ref.~\cite{Hansen:2019idp} to determine an inclusive rate in the $1+1$ dimensional O(3) non-linear $\sigma$-model, analogous to the QCD part of ${e}^+{e}^- \rightarrow \rm {hadrons}$. The Euclidean two-point correlation function of the conserved current $j$ is computed using Monte Carlo lattice field theory simulations for a variety of spacetime volumes and lattice spacings. The spectral density of this correlator is related to the inclusive rate for $j \rightarrow {\rm X}$ in which all final states produced by the external current are summed. The ill-posed inverse problem of determining the spectral density from the correlation function is made tractable through the determination of smeared spectral densities in which the desired density is convolved with a set of known smearing kernels of finite width $\epsilon$. The smooth energy dependence of the underlying spectral density enables a controlled $\epsilon \to 0$ extrapolation in the inelastic region, yielding the real-time inclusive rate without reference to individual finite-volume energies or matrix elements. Systematic uncertainties due cutoff effects and residual finite-volume effects are estimated and taken into account in the final error budget. After taking the continuum limit, the results are consistent with the known analytic rate to within the combined statistical and systematic errors. Above energies where 20-particle states contribute, the overall precision is sufficient to discern the four-particle contribution to the spectral density.
\end{abstract}

\nopagebreak
\maketitle

\vfill

\pagebreak

\tableofcontents

\vfill
\pagebreak

\section{Introduction}\label{s:intro}

Markov chain Monte Carlo simulations of lattice QCD continue to provide a quantitative window into the strong nuclear force. However, the Euclidean metric signature required for Monte Carlo sampling of the QCD path integral complicates the study of real-time scattering processes. While the spatial volume dependence of energies and matrix elements has successfully been used as a probe of few-particle scattering and decay amplitudes~\cite{Luscher:1990ux,Lellouch:2000pv}, this approach is inapplicable to energies above arbitrary multi-particle thresholds\footnote{For recent reviews on the status of treating three-particle states see Refs.~\cite{Hansen:2019nir,Rusetsky:2019gyk}.} and is restricted to the discrete set of center-of-mass energies that appear in a given finite-volume geometry. This last disadvantage hampers the continuum limit of amplitudes at fixed energy, which can only be achieved by maintaining a constant physical volume as the lattice spacing is decreased. Using the finite-volume formalism to determine inclusive rates is also difficult as these can only be calculated by extracting all contributing amplitudes separately and summing the individual rates.

An alternative approach, following the seminal work of Ref.~\cite{Barata:1990rn}, is to use Euclidean correlation functions computed in lattice simulations to determine spectral densities which are definitionally independent of the metric signature. This alternative is not subject to the same limitations as finite-volume methods and has been formally developed for inclusive rates mediated by an external current~\cite{Liu:2016djw,Hansen:2017mnd,Gambino:2020crt} as well as arbitrary scattering and transition amplitudes~\cite{Bulava:2019kbi, Bruno:2020kyl}.\footnote{This approach is inspired in part by the extraction of analogous observables in lattice calculations at finite temperature. See Ref.~\cite{Meyer:2011gj} for an instructive review.} A major obstacle of this program is the solution of an ill-posed inverse problem to determine continuous spectral densities from correlation functions sampled on a finite set of time separations with statistical errors. In Refs.~\cite{BG1,BG2} Backus and Gilbert deal with this by instead computing spectral densities smeared with a kernel that is known \emph{a posteriori}. An important advancement by Ref.~\cite{Hansen:2019idp} (see also Refs.~\cite{pt_hlt,Bailas:2020qmv}) enables the smearing kernel to be specified \emph{a priori}. A review and comparison of alternative spectral reconstruction techniques goes beyond the scope of this work~\cite{Maass1990_0e27e,Tripolt:2018xeo}.

Although these methods are not new, their application to compute inclusive rates and scattering amplitudes from actual Monte Carlo simulation data, with a detailed analysis of both statistical and systematic errors, has not yet been performed. It is therefore worthwhile to pursue such a test in a controlled setting. This work employs the spectral reconstruction strategy of Ref.~\cite{Hansen:2019idp} to determine an inclusive rate in the two-dimensional O(3) non-linear $\sigma$-model. As in Ref.~\cite{Luscher:1990ck}, comparison is made with exact continuum results. However, our work overcomes the limitations mentioned above: the inclusive rate is computed above inelastic thresholds and the continuum limit taken at fixed center-of-mass energy without the need for a constant physical volume. Large physical simulation volumes are however required to control finite-volume effects in the smeared spectral density at fixed smearing width. After taking the continuum limit and accounting for systematic errors due to finite-volume effects, the desired spectral density is obtained by extrapolating the smearing width to zero.

For this first application we treat a process akin to ${e}^+ {e}^- \rightarrow {\rm hadrons}$, the QCD component of which is given by the spectral density
\begin{align}
\label{eq:rhomunu}
\rho_{\mu\nu}(k) = \frac{1}{2\pi}\int d^4 x \, e^{-ik\cdot x} \langle
\Omega | \hat{j}^{\rm em}_{\mu}(x)\, \hat{j}^{\rm em}_{\nu}(0) | \Omega \rangle = (g_{\mu\nu}k^2-k_\mu k_{\nu}) \, \rho(k^2),
\end{align}
where the integral is performed in Minkowski space, $\hat{j}^{\rm em}_\mu$ is the quark-level electromagnetic current and the relation to the physical process is
\begin{align}
\rho(s) = \frac{R(s)}{12\pi^2}, \quad R(s) = \frac{\sigma\left[{e}^+ {e}^- \rightarrow {\rm hadrons}\right](s)}{4\pi\alpha_{\rm em}(s)^2/(3s)}.
\end{align}
This celebrated `$R$-ratio' has a number of phenomenological applications including the hadronic vacuum polarization contribution to the anomalous magnetic moment of the muon. It can be obtained via spectral reconstruction of the Euclidean correlator in the time-momentum representation
\begin{align}\label{e:qcd_tmr}
C(t) =  \int d^3 \boldsymbol{x} \, \langle \Omega | \hat{j}^{\rm em}_z(\boldsymbol{x}) \, e^{-\hat{H}t} \, \hat{j}^{\rm em}_z(0)^\dagger | \Omega \rangle = \int_0^{\infty} d\omega \, \omega^2 \rho(\omega^2) \, e^{-\omega t} \,,
\end{align}
computed in lattice QCD simulations.\footnote{As written, this expression holds for both the anti-hermitian definition of $\hat{j}^{\rm em}_z(0)$, conventionally used with the Euclidean metric, as well as the hermitian Minkowski current implicitly used in Eq.~\eqref{eq:rhomunu}.} Everything discussed in this work is readily transferable to the determination of $\rho(s)$ from $C(t)$ in lattice QCD. While computations of $\rho(s)$ are naturally compared with experimental determinations of $R(s)$, numerical results for the spectral density computed here are compared with the corresponding analytical predictions in Fig.~\ref{f:inel}, which is our main result.

This remainder of this work is organized as follows. Sec.~\ref{s:exact} defines the spectral density of interest in the O(3) model, outlines the reconstruction approach, and suggests a strategy for extrapolating the smearing width to zero. Sec.~\ref{s:fv} examines the influence of the finite torus on which the simulations are performed while Sec.~\ref{s:lat} defines the lattice regularization and discusses the continuum limit. Sec.~\ref{s:res} presents numerical results and Sec.~\ref{s:conc} concludes.

\section{General framework in continuous infinite volume}\label{s:exact}

This section introduces the two-dimensional O(3) non-linear $\sigma$-model and the Euclidean correlator $C(t)$ with spectral density $\rho(\omega)$, which are both defined via the conserved vector current. We additionally review the algorithm of Ref.~\cite{Hansen:2019idp} for systematically determining a smeared version of $\rho(\omega)$ from numerical estimates of $C(t)$ and detail an extrapolation procedure for taking the smearing width to zero. In this section the Euclidean spacetime is assumed to be continuous and infinite in both directions. Peculiarities due to the finite simulation volume are discussed in Sec.~\ref{s:fv} and the lattice discretization in Sec.~\ref{s:lat}.

The continuum Euclidean action of the two-dimensional O(3) model is defined as
\begin{align}\label{e:act}
S[\sigma] = \frac{1}{2g^2} \int d^2x \, \partial_{\mu} \sigma(x) \cdot \partial_{\mu} \sigma(x) \,,
\end{align}
where the $3$-component real field $\sigma^a(x)$ has unit length $\sigma(x) \cdot \sigma(x) = 1$. The O(3) model is asymptotically free, has a dynamically-generated mass gap $m$, and is integrable. It also possesses a global O(3) symmetry which rotates the field $\sigma(x)$. The corresponding Noether current is given by
\begin{align}
j_{\mu}^{c}(x) = \frac{1}{g^2} \epsilon^{abc} \sigma^a(x) \partial_{\mu} \sigma^b(x) \,,
\end{align}
where $\epsilon^{abc}$ is the Levi-Civita tensor and repeated indices are summed. This current transforms irreducibly under the $I=1$ (fundamental) representation of O(3).

The aim of this work is to reconstruct the spectral density $\rho(\omega)$ associated with $j_{\mu}^{c}$ using Euclidean correlation functions determined numerically from lattice Monte Carlo simulations. The spectral density is defined implicitly via the relation
\begin{gather}\label{e:rfact}
2 \pi \, \langle \Omega \vert \hat{j}^a_\mu(0) \, \delta^2(\hat{P}-p) \, \hat{j}^b_\nu(0) \vert \Omega \rangle
=
\frac{\delta^{ab}}{3} \left( \delta_{\mu\nu} - \frac{p_\mu p_\nu}{p^2} \right) \rho\left(\sqrt{p^2}\right)
\, ,
\end{gather}
where $\hat{P} = (\hat{H},\hat{\boldsymbol{P}})$ is the 2-momentum operator and $p$ is an external 2-momentum coordinate with components denoted $p = (E, \boldsymbol{p})$. Covariance under both the O(3) internal symmetry and the Euclidean spacetime SO(2) symmetry, as well as current conservation, enable the factorization in Eq.~\eqref{e:rfact}. By specializing $\mu=\nu=1$ and $p=(E,0)$, and contracting the internal indices, one obtains the more direct definition
\begin{gather}\label{e:rdef}
\rho(E)
=
 2 \pi \, \langle \Omega \vert \hat{j}^a_1(0) \, \delta^2(\hat{P}-p) \, \hat{j}^a_1(0) \vert \Omega \rangle
\, .
\end{gather}

The utility of the two-dimensional O(3) model for the present study is that (due to integrability) $\rho(E)$ can be computed analytically, enabling a comparison between our numerical reconstruction and the exact result. The spectral density decomposes into sectors defined by the number $n$ of asymptotic particles\footnote{The $n$-particle states interpolated by the current also transform irreducibly under the $I=1$ representation of O(3).} propagating between the two currents
\begin{equation}
\rho(E) = \sum_{\text{even } n \ge 2} \rho^{(n)}(E) \,.
\end{equation}
The contribution $ \rho^{(n)}(E) $ has support for $E > n\, m$, where $m$ is the mass gap. Even though integrability (together with a number of mild and generally accepted assumptions) fixes $\rho^{(n)}(E)$ for every even value of $n$, explicit expressions have been worked out only for $n=2,4,6$. At the energies considered here the sum over the number of particles is rapidly convergent and the $n=6$ contribution is at least a couple of orders of magnitude smaller than the $n=4$ contribution. In the following we therefore refer to the spectral density summed over $n=2,4,6$ as the \textit{exact spectral density}. The $n=2$ contribution can be written in closed form
\begin{gather}\label{e:2part}
\rho^{(2)}(E) =
\frac{3\pi^3}{8\theta^2}
\,
\frac{\theta^2 + \pi^2}{\theta^2 + 4\pi^2} \,
\tanh^3 \frac{\theta}{2} \, \, \bigg \vert_{\theta = 2\cosh^{-1} \frac{E}{2m}} \, ,
\end{gather}
while the $n=4,6$ contributions are expressed in terms of phase-space integrals that require numerical evaluation as discussed in App.~\ref{s:app_sd}.

This spectral density is related to the Euclidean-signature current-current correlator at zero spatial momentum via the Laplace transform
\begin{align}\label{e:corr}
C(t) \equiv \int d\boldsymbol{x} \, \langle \Omega | \, \hat{j}_1^{a}(0, \boldsymbol{x}) \, e^{-\hat{H}t } \,
\hat{j}_1^{a}(0) \, | \Omega \rangle
=
\int_0^{\infty}d\omega \, e^{-\omega t} \, \rho(\omega)
\, .
\end{align}
The demonstration of a systematic method to invert this relation, given realistic numerical estimates of $C(t)$ at a finite set of time slices with statistical errors, is the central focus of this work.

A ubiquitous problem, faced across a breadth of scientific disciplines, is the numerically ill-conditioned nature of the inverse Laplace transform.\footnote{See for example Ref.~\cite{Maass1990_0e27e} and references therein.} A promising way forward is to recognize that $\rho(\omega)$ is not directly extractable from numerical data and that one should instead target the smeared spectral density
\begin{align}
\label{e:rhosmear}
 \rho_{\epsilon}(E) = \int_0^\infty d\omega \, \delta_{\epsilon}(E, \omega) \, \rho(\omega) \,,
\end{align}
where $\delta_{\epsilon}(E, \omega)$ is any approximation of the Dirac $\delta$-function satisfying $\lim_{\epsilon \to 0} \delta_{\epsilon}(E, \omega) = \delta(E - \omega)$ and $\int_{-\infty}^\infty d \omega\, \delta_{\epsilon}(E, \omega) = 1$. The challenge of recovering $\rho_{\epsilon}(E)$ from $C(t) $ can then be made arbitrarily mild (or severe) by varying the specific functional form of $\delta_{\epsilon}(E, \omega) $ and the values of $\epsilon$ and $E$.

Following Ref.~\cite{Hansen:2019idp} we consider smearing kernels $\delta_{\epsilon}(E, \omega)$ that can be represented exactly as
\begin{align}\label{e:delta}
 {\delta}_{\epsilon}(E,\omega) = a\sum_{\tau
 = 1}^{\infty} g_\tau^{\rm target} \, b_{\tau}(\omega) \,,
\end{align}
where $\tau$ is a dimensionless integer variable, $a$ an arbitrary scale with dimensions of inverse energy to be later identified with the lattice spacing, the $b_{\tau}(\omega)$ are basis functions, and $g_\tau^{\rm target}\equiv g_\tau^{\rm target}(\epsilon,E)$ components of the target smearing kernel ${\delta}_{\epsilon}(E,\omega)$ in this basis. If we identify $b_\tau(\omega) = {\rm e}^{-a\omega \tau}$, then the smeared spectral density is given by
\begin{equation}
{\rho}_{\epsilon}(E) = a\sum_{\tau = 1}^{\infty} g^{\rm target}_{\tau} C(a\tau) = \int_0^\infty d\omega \, \delta_{\epsilon}(E, \omega) \, \rho(\omega) \,.
\end{equation}
The choice $b_\tau(\omega) = {\rm e}^{-a\omega \tau}$ is based on the non-essential assumption that spacetime is infinite. In practice, as explained in Sec.~\ref{s:fv}, we use a basis that takes into account the periodicity of the finite temporal direction.

An estimator for $\rho_{\epsilon}(E)$ is obtained by approximating $\delta_{\epsilon}(E, \omega)$ with an element of the space spanned by a finite number of basis functions. The coefficients $g_\tau$ corresponding to the approximation of $\delta_{\epsilon}(E, \omega)$ are determined by minimizing the functional\footnote{Possible additional constraints on the minimization are discussed in App.~\ref{a:hlt}.}
\begin{align}\label{e:Wfunc}
W_{\lambda}[g] = (1-\lambda)\frac{A[g]}{A[0]} + \lambda B[g] \,,
\end{align}
where $\lambda\in[0,1]$ is the `trade-off' parameter (discussed shortly) and the functionals $A[g]$ and $B[g]$ are given by\footnote{The relative normalization of the $A[g]$ and $B[g]$ functionals in Eq.~\eqref{e:Wfunc} differs from Ref.~\cite{Hansen:2019idp}. The factor $1/(aC(0))^2$ in the normalization of $B[g]$ present there is not included here since $aC(0)\simeq 1$ while the factor $1/A[0]$ in the first term of Eq.~\eqref{e:Wfunc} makes it dimensionless ($A[0]\propto 1/\epsilon$).}
\begin{align}
A[g] =
\int_{E_0}^\infty d\omega \left\{ {\delta}_{\epsilon}(E,\omega) - a\sum_{\tau =1}^{\tau_{\rm max}} g_\tau \, b_{\tau}(\omega)\right\}^2\,,
\qquad
B[g] = \sum_{\tau, \tau^\prime=1}^{\tau_{\rm max}} g_{\tau} g_{\tau^\prime} \, {\rm Cov} \left[aC(a\tau), aC(a\tau^\prime) \right] \,.
\label{eq:AandB}
\end{align}
As discussed in App.~\ref{a:hlt}, $A[g]$ and $B[g]$ are both positive quadratic forms so that the minimum conditions
\begin{align}
\left.\frac{\partial W_{\lambda}[g]}{\partial g_\tau}\right\vert_{g_\tau=g_\tau^\lambda}=0 \,,
\label{eq:Wminimumg}
\end{align}
are a linear system of equations to be solved for the coefficients $g_\tau^\lambda\equiv g_\tau^\lambda(\epsilon,E,\tau_{\rm max})$. These coefficients define the approximation of $\delta_{\epsilon}(E, \omega)$ and the associated estimator for ${\rho}_{\epsilon}(E)$ according to
\begin{align}
{\delta}^\lambda_{\epsilon}(E,\omega) = a\sum_{\tau
 = 1}^{\tau_{\rm max}} g^\lambda_\tau \, b_{\tau}(\omega)\;,
\qquad
{\rho}^\lambda_{\epsilon}(E) = a\sum_{\tau = 1}^{\tau_{\rm max}} g^\lambda_{\tau} C(a\tau)
= \int_0^\infty d\omega \, \delta^\lambda_{\epsilon}(E, \omega) \, \rho(\omega)\;.
\end{align}
The functional $B[g^\lambda]$ is simply the statistical variance of $\rho^\lambda_{\epsilon}(E)$ and therefore vanishes in the ideal case of infinitely precise input data. On the other hand, $A[g^\lambda]$ measures the distance between the target kernel ${\delta}_{\epsilon}(E,\omega)$ and its approximation $\delta^\lambda_{\epsilon}(E, \omega)$ in the range\footnote{The parameter $E_0$ can be adjusted by exploiting the fact that $\rho(E)$ has support only for $E> 2 m$, so that $\rho_\epsilon(E)$ in Eq.~\eqref{e:rhosmear} is insensitive to $\delta_{\epsilon}(E, \omega)$ for $\omega < 2 m$. The same holds for $\rho^\lambda_\epsilon(E)$ so that the functional form of $\delta^\lambda_{\epsilon}(E, \omega)$ can be left unconstrained for $\omega < 2 m$. Any $E_0 \leq 2 m$ is therefore a viable choice in determining the coefficients $g_\tau^\lambda$ so $E_0$ can be chosen to improve the numerical stability of the minimization procedure.} $E\in [E_0,\infty]$. It can only vanish in the limit $\tau_{\rm max}\rightarrow \infty$ when $g^\lambda_\tau \rightarrow g^{\rm target}_{\tau}$. The coefficients $g^\lambda_\tau$ that minimize $W_{\lambda}[g]$ thus represent a particular balance between statistical and systematic errors, as dictated by the $\lambda$ parameter. For small $\lambda$ the estimator $\rho^\lambda_{\epsilon}(E)$ is close to $\rho_{\epsilon}(E)$ but with a large statistical uncertainty. Conversely, for large $\lambda$ the estimator $\rho^\lambda_{\epsilon}(E)$ has a small statistical error but differs significantly from $\rho_{\epsilon}(E)$.

When evaluated at the minimum, the functional $W_\lambda[g]$ is a function of $\lambda$ only, thus defining $W(\lambda)\equiv W_\lambda[g^\lambda]$. The recipe suggested in Ref.~\cite{Hansen:2019idp} to choose the optimal value of the trade-off parameter defines $\lambda_\star$ such that
\begin{align}
\left. \frac{\partial W(\lambda)}{\partial \lambda}\right\vert_{\lambda=\lambda_\star} =0\;.
\label{eq:Wmax}
\end{align}
A straightforward application of Eq.~\eqref{eq:Wminimumg} demonstrates that at $\lambda_\star$ (the maximum of $W(\lambda)$ where $g_\tau^{\lambda}=g_\tau^\star$) one has $A[g^\star]=A[0]B[g^\star]$. This can be understood as the condition of `optimal balance' between statistical and systematic errors. The numerical results presented here are obtained using this recipe. In the remainder of this work, unless explicitly stated, we do not distinguish the theoretical quantity $\rho_\epsilon(E)$ from its numerical estimator $\rho^{\lambda_\star}_{\epsilon}(E)$. Estimates of the systematic error on $\rho_\epsilon(E)$ induced by the residual difference $\delta^{\lambda_\star}_{\epsilon}(E,\omega)-\delta_{\epsilon}(E,\omega)$ are discussed in Sec.~\ref{s:res}.

A defining feature of the approach in Ref.~\cite{Hansen:2019idp} is that the smearing kernel $\delta_{\epsilon} (E, \omega)$ and the associated values of $\epsilon$ and $E$ are inputs of the algorithm, in contrast to the original Backus-Gilbert method~\cite{BG1, BG2}. This work exploits this by employing four functional forms for the smearing kernel, each of which is a function of $x = E - \omega$:
\begin{align}\label{e:kers}
\delta_{\epsilon}^{\sf g}(x) & = \frac{1}{\sqrt{2\pi}\epsilon} \exp\left[-\frac{x^2}{2\epsilon^2}\right], &
\delta_{\epsilon}^{{\sf c}0}(x) & = \frac{1}{\pi} \frac{\epsilon}{x^2 + \epsilon^2} \,, \\[5pt]
\delta_{\epsilon}^{{\sf c}1}(x) & = \frac{2}{\pi} \frac{ \epsilon ^3}{ ( x^2 + \epsilon ^2 )^2} \,,
 & \delta_{\epsilon}^{{\sf c}2}(x) & = \frac{8}{3 \pi} \frac{ \epsilon ^5}{ ( x^2 + \epsilon ^2 )^3}
\,.
\end{align}
Here ${\sf g}$ and ${\sf c}$ stand for `Gauss' and `Cauchy' respectively, and the number following ${\sf c}$ gives the order of the Cauchy-like pole as shown. All kernels are normalized to unit area.

\begin{table}
\renewcommand{\arraystretch}{1.7}
\begin{tabular}{c c c || c c c c}
\toprule
\ \ \ {\sf x} \ \ \ & \ \ \ $w^{{\sf x}}_k,$ even\ $k$ \ \ \ & \ \ \ $w^{{\sf x}}_k,$ odd\ $k$ \ \ \ & \ \ \ $\phantom{+} w^{{\sf x}}_1$ \ \ \ & \ \ \ $\phantom{+} w^{{\sf x}}_2$ \ \ \ & \ \ \ $\phantom{+} w^{{\sf x}}_3$ \ \ \ & \ \ \ $\phantom{+} w^{{\sf x}}_4$ \ \ \
\\ \midrule
{\sf g} & $\displaystyle \frac{k!}{(-2)^{k/2} (k/2)!}$ & $0$ & $\phantom{+} 0$ & $-1$ & $\phantom{+} 0$ & $\phantom{+} 3$ \\
{\sf c0} & $\displaystyle \phantom{\frac{1}{2}} 1 \phantom{\frac{1}{2}} $ & $\displaystyle \phantom{\frac{1}{2}} 1 \phantom{\frac{1}{2}} $ & $\phantom{+} 1$ & $\phantom{+} 1$ & $\phantom{+} 1$ & $\phantom{+} 1$
\\
{\sf c1} & $\phantom{\frac{1}{2}} (1-k) \phantom{\frac{1}{2}}$ & \ \ $(1-k)$ \ \ & $\phantom{+} 0$ & $-1$ & $-2$ & $-3$
\\
{\sf c2} & \ \ $\displaystyle \frac{1}{3}(k-3)(k-1)$ \ \ & \ $\displaystyle \frac{1}{3}(k-3)(k-1)$ \ & $\phantom{+} 0$ & $-1/3$ & $\phantom{+} 0$ & $\phantom{+}1$ \\
\bottomrule
\end{tabular}
\caption{Summary of the kernel-dependent coefficients $w^{{\sf x}}_k$ entering the small-$\epsilon$ expansion in Eq.~\eqref{e:expand}. Although not reflected in the general formulae, $w_3^{{\sf c1}}$ and $w_5^{{\sf c2}}$ are the lowest non-zero coefficients with odd $k$ for their respective smearing kernels. \label{t:wcoeffs}}
\renewcommand{\arraystretch}{1.0}
\end{table}

Given estimates of $\rho_{\epsilon}(E)$ over a range of $\epsilon$ for each of the resolution functions shown above, the final step in determining $\rho(E)$ is to perform an $\epsilon \to 0$ extrapolation. To this end it is useful to understand the small-$\epsilon$ expansion at fixed $E$. We consider this expansion only at energies away from singularities, i.e.~away from $2 m \mathbb Z^+$. At such points the smeared spectral density satisfies
\begin{align}
\label{e:expand}
\rho^{{\sf{x}}}_{\epsilon}(E) & \equiv \int_0^\infty d \omega \, \delta_{\epsilon}^{\sf x}(E - \omega) \, \rho(\omega) = \rho(E) + \sum_{k=1}^\infty w^{{\sf x}}_k a_k(E) \epsilon^{k} \,,
\end{align}
where the superscript $\sf x$ labels a particular smearing kernel. As indicated in the rightmost expression, the $O(\epsilon^k)$ contribution to the expansion factorizes into a geometric kernel-dependent coefficient $w^{{\sf x}}_k$ and a kernel-independent piece $a_k(E)$ which depends on $\rho(E)$ only. Tab.~\ref{t:wcoeffs} summarizes the values of $w^{{\sf x}}_k$ for the four kernels used in this work. The ambiguity in separating $w^{{\sf x}}_k$ and $a_k(E)$ is fixed by setting $w^{{\sf c0}}_k = 1$ for all $k$. For this choice, all remaining $w^{{\sf x}}_k$ are rational numbers and
\begin{equation}
\label{e:ak}
a_{k}(E) =
\begin{cases}
\frac{(-1)^{k/2}}{k!} \left( \frac{d}{dE} \right)^{k}\rho(E)\,,
\qquad & \textrm {$k$ even} 
\\
\\
\lim_{\eta \to 0^+} \frac{(-1)^{(k-1)/2}}{2\pi} \int_{-\infty}^\infty {\rm d}\omega \, \frac{\rho(E+\omega) + \rho(E-\omega)}{(\omega+i\eta)^{k+1}}\,,
\qquad & \textrm{$k$ odd}
\end{cases}
\ .
\end{equation}

\section{Finite-volume estimator}\label{s:fv}

The theory is considered here on an $L \times T$ volume with periodic boundary conditions in both directions. Finite-$L$ and finite-$T$ effects on the spectral density are significantly different. On the one hand, finite-$T$ (or thermal) effects are shown below to be exponentially suppressed and are therefore reliably small. On the other hand, the spectral density is dramatically different at finite and infinite $L$.

At infinite $L$ the spectral density $\rho(E)$ defined in Eq.~\eqref{e:rdef} is a continuous function which is analytic for all $E$ except for $E=n\,m$ where $n$ is a positive even integer. At finite $L$, by contrast, the spectrum of the Hamiltonian is discrete and the spectral density is a sum of Dirac $\delta$-functions. As $L$ increases, the spectrum becomes denser and denser so that the continuous $L=\infty$ spectral density is recovered as a weak (or distributional) limit. In no meaningful way can this limit be considered point-wise and the finite-$L$ effects on $\rho(E)$ treated as small corrections. Questions such as ``Are the finite-$L$ corrections to $\rho(E)$ exponentially suppressed?'' are simply ill-posed. By contrast, smeared spectral densities converge to their infinite-$L$ value in a point-wise sense. An immediate consequence of this discussion is that the $L \to \infty$ and $\epsilon \to 0^+$ limits (where $\epsilon$ is the width of the smearing kernel) do not commute and make sense only in a precise order: the $L \to \infty$ limit must be taken before the $\epsilon \to 0^+$ limit. The importance of the ordered double limit was also stressed in Ref.~\cite{Hansen:2017mnd} where the combined $L$ and $\epsilon$ dependence of related spectral functions was considered in a perturbative context.

We know that individual finite-$L$ matrix elements at energies above the two-particle threshold approach their infinite-$L$ limit with corrections which vanish as inverse powers of $1/L$. When smeared spectral densities are considered, one may hope that finite-$L$ corrections vanish faster than any inverse power in $1/L$, at least if the smearing kernel is `reasonable enough'. This issue is explored in App.~\ref{s:app_fv}, which considers a fictitious system where the infinite-$L$ spectral density is given entirely by the two-particle contribution in Eq.~\eqref{e:2part}. The finite-$L$ spectral density is then determined (up to exponential corrections) by the Lellouch-L\"uscher formalism~\cite{Lellouch:2000pv, Meyer:2011um}. One sees explicitly that the smeared spectral density has finite-$L$ contributions which are $O({\rm e}^{-mL})$ in the case of the Gaussian kernel, despite the power-law corrections to the individual energies and matrix elements. For the Cauchy kernels the effects are also exponentially suppressed by $O({\rm e}^{- \mu L})$ where $\mu \leq m$ depends on the values of $E, \epsilon$ and $m$ as specified in Eq.~\eqref{e:mudef}.

In the derivation of the exponentially-suppressed finite-$L$ contributions, the analyticity of the smearing kernel plays a central role. Smooth but non-analytic kernels produce finite-$L$ corrections that decay faster than any inverse power of $L$ but generally not exponentially. We also observe that the finite-volume effects are oscillatory functions of $L$ with an envelope decaying according to the predicted scaling. We do not claim that these findings are valid beyond the simple exercise considered here, but it is clear that the landscape of phenomena related to the $L \to \infty$ limit of smeared spectral densities is rich.

After this lengthy introduction, we give some explicit formulae. At finite $L$ and $T$, the zero-momentum current-current correlator has the following Hamiltonian representation
\begin{gather}
C_{T,L}(t)
=
\frac{1}{L}
\frac{
 \text{tr}\, \{ {\rm e}^{-(T-t)\hat{H}_L} \hat{A} {\rm e}^{-t \hat{H}_L} \hat{A} \}
}{
 \text{tr}\, {\rm e}^{-T\hat{H}_L}
}
\ ,
\end{gather}
where the definition
\begin{gather}
\hat{A} = \int_0^L d\boldsymbol{x} \, \hat{j}_1(\boldsymbol{x}) \,,
\end{gather}
is employed. At finite $L$, the spectrum of the Hamiltonian is discrete. By introducing an orthonormal basis of energy eigenstates $| n \rangle_L$ satisfying $\hat{H}_L | n \rangle_L = E_n(L) | n \rangle_L$, one easily derives the spectral representation of the correlator
\begin{gather}
C_{T,L}(t)
=
\int_{-\infty}^\infty d\omega \, \tilde{\rho}_{T,L}(\omega) e^{-t \omega}
\, ,
\label{eq:CTL_rho2}
\end{gather}
with the definition
\begin{gather}
\tilde{\rho}_{T,L}(\omega) = \frac{1}{L} \sum_{n,n'} \frac{ e^{-TE_{n'}(L)} }{ \sum_{n''} e^{-TE_{n''}(L)} }
\left| \langle n' | \hat{A} | n \rangle_L \right|^2
\delta( E_n(L)-E_{n'}(L) - \omega )
\, .
\label{eq:rhoTL1}
\end{gather}
Notice that $\tilde{\rho}_{T,L}(\omega)$ is non-vanishing also for negative values of $\omega$. By separating terms with $E_n \ge E_{n'}$ and $E_n \le E_{n'}$, and taking care to avoid double counting contributions with $E_n = E_{n'}$, one can split $\tilde{\rho}_{T,L}(\omega)$ into
\begin{gather}
\tilde{\rho}_{T,L}(\omega) = \rho_{T,L}(\omega) + e^{T\omega} \rho_{T,L}(-\omega)
\ ,
\label{eq:rhoTL2}
\end{gather}
where the spectral density
\begin{gather}
\rho_{T,L}(E)
=
\frac{1}{L} \sum_{E_n \ge E_{n'}} \frac{ e^{-TE_{n'}(L)} }{ \sum_{n''} e^{-TE_{n''}(L)} }
\left( 1 - \frac{\delta_{E_n,E_{n'}}}{2} \right) \left| \langle n' | \hat{A} | n \rangle_L \right|^2
\delta( E_n(L)-E_{n'}(L) - E )
\, ,
\label{eq:rhoTL3}
\end{gather}
vanishes for $E < 0$. By plugging this information in the correlator, we obtain the spectral representation in the form
\begin{gather}
C_{T,L}(t)
=
\int_{0^-}^\infty dE \, \rho_{T,L}(E) \{ e^{-tE} + e^{-(T-t)E} \}
\, .
\label{eq:CTL_rho2}
\end{gather}

As described in Sec.~\ref{s:exact}, for a target smearing kernel $\delta_\epsilon(E-\omega)$, one now seeks an approximation $\delta^{\lambda}_{\epsilon}(E,\omega)$ in the space generated by the function basis
\begin{gather}
b_{T,\tau}(\omega) = e^{-a\tau\omega} + e^{-(T-a\tau)\omega}
\,,
\end{gather}
by minimizing the functional $W_{\lambda}[g]$. Then
\begin{gather}
{\rho}^{\lambda}_{T,L,\epsilon}(E)
=
 a\sum_{\tau = 1}^{\tau_{\rm max}} g^{\lambda}_\tau C_{T,L}(a\tau)
=
\int_{0^-}^\infty dE \, \rho_{T,L}(E) {\delta}^{\lambda}_{\epsilon}(E,\omega)
\label{eq:estimator_with_TL}
\end{gather}
is an approximation for the smeared spectral density
\begin{gather}
{\rho}_{T,L,\epsilon}(E)
=
\int_{0^-}^\infty dE \, \rho_{T,L}(E) {\delta}_{\epsilon}(E-\omega)
\, .
\end{gather}
A couple of comments on these formulae are in order. \textit{(1)} The estimator in Eq.~\eqref{eq:estimator_with_TL} formally depends on $\tau_{\rm max}$, but (as demonstrated in Sec.~\ref{s:res}) our results are rather insensitive to its particular value. \textit{(2)} The coefficients $g_\tau^\lambda$ obtained by solving the minimization problem in Eq.~\eqref{eq:Wminimumg} depend on $L$ and $T$ via the covariance matrix appearing in $B[g]$, and on $T$ via our choice of the basis $b_{T,\tau}(\omega)$. In the limit of infinite statistics (with $B[g]=0$), the $g_\tau^\lambda$ depend on $T$ but not on $L$.

Eq.~\eqref{eq:rhoTL3} makes manifest that finite-temperature effects on the spectral density at fixed $L$ are exponentially suppressed, i.e.
\begin{gather}
\rho_{T,L}(E) =
\rho_{\infty,L}(E)
+ O({\rm e}^{-Tm(L)})
=
\frac{1}{L} \sum_{n}
\left| \langle 0 | \hat{A} | n \rangle_L \right|^2
\delta( E_n(L) - E )
+ O({\rm e}^{-Tm(L)})
\, ,
\label{eq:rhoTL-Tinf}
\end{gather}
where $m(L)$ is the energy gap in finite volume. Smearing the spectral density amounts to replacing the Dirac $\delta$-function in Eq.~\eqref{eq:rhoTL3} with the smearing kernel $\delta_{\epsilon}$. At fixed $L$ the smeared spectral density also has finite-temperature effects that are exponentially suppressed with $O({\rm e}^{-Tm(L)})$. Even though far from obvious, it is reasonable to assume that finite-$T$ effects are exponentially suppressed also at $L=\infty$ since the theory has a mass gap.

\section{Lattice-discretized model}\label{s:lat}

We calculate the Euclidean current-current correlator in Eq.~\eqref{eq:CTL_rho2} by numerical simulation of the lattice-discretized model. The field $\sigma(x)$ is now defined on the set of $(L/a)\times(T/a)$ equally-spaced points $x = (a\tau, \boldsymbol{x})$ on the two-torus, denoted by $\Lambda$. The standard discretization of the action is employed here
\begin{align}\label{e:slat}
S[\sigma] = \frac{\beta}{2} \sum_{x \in \Lambda} a^2
\sum_\mu \hat{\partial}_{\mu} \sigma(x) \cdot \hat{\partial}_{\mu} \sigma(x)
=
\beta \sum_{x \in \Lambda}
\sum_\mu \left[ 1 - \sigma(x) \cdot \sigma(x + a\hat{\mu}) \right]
\,,
\end{align}
where $\hat{\partial}_{\mu} f(x) = \frac{1}{a} [f(x + a\hat{\mu}) - f(x)]$ is the forward-difference operator defined with periodic boundary conditions in both directions. Expectation values of observables are given by the path integral
\begin{align}\label{e:path}
\langle A \rangle = \frac{1}{Z} \int \left[ \prod_{x \in \Lambda}
{\rm d}\sigma(x) \right] \, {\rm e}^{-S[\sigma]} A[\sigma] \,,
\end{align}
where ${\rm d}\sigma(x)$ denotes the O(3)-invariant integration measure over the unit sphere. The global O(3) symmetry is preserved by the lattice discretization, and its associated Noether current is
\begin{align}\label{e:lat_cur}
j_{\mu}^a(x) = \beta \epsilon^{abc} \sigma^b(x) \hat{\partial}_{\mu} \sigma^c(x)
\,.
\end{align}
Exact O(3) Ward identities imply that this current is renormalized. The discretized version of the zero-momentum Euclidean current-current correlator is given by
\begin{align}
C(a\tau) = \sum_{\boldsymbol{x}} a \langle j_{1}^a(x) \,
j_{1}^a(0) \rangle \,.
\end{align}

The strong dynamics of the O(3) model generate a mass gap $m$, which is often used to set the scale. However, in this work we determine all dimensionful quantities in units of the appropriate power of the alternative scale $m_\star$, defined as~\cite{Francis:2013jfa}
\begin{gather}\label{e:m}
m_\star^{-1} C(m_\star^{-1}) = 0.046615 \, .
\end{gather}
While in principle one can choose any number in the right-hand side, we use a value that gives $m_\star \simeq m$. Using the correlator reconstructed from the $2$-, $4$- and $6$-particle contributions to the spectral density we find that the relative difference between $m$ and $m_\star$ is of order $10^{-5}$. In practice the correlator $C(t)$ is known only at values of $t=a\tau$ that are integer multiples of the lattice spacing $a$, and the equation for $m_\star$ is solved using a piecewise linear interpolation of $\log t C(t)$. The quantity $m_\star$ is determined with higher statistical precision than $m$ and is also affected by smaller finite-volume effects. It furthermore does not require any additional correlation functions. 

\begin{table}
\begin{tabular}{ccccccccccc}
\toprule
ID & $(L/a)\times(T/a)$ & $\beta$ & $am_\star$ & $m_\star L$ & $m_\star T$ & $N_{\rm th}\times 10^{-6}$ & $N_{\rm rep}$ & $N_{\rm bin}$ & $B\times 10^{-6}$ & $N_{\rm c}\times 10^{-10}$ \\
\midrule
A1 & $640 \times 320 $ & 1.63 & 0.0447989(62) & 29 & 14 & 1.6 & 480 & 5 & 3.2 & 3.84\\ %
A2 & $1280 \times 640 $ & 1.72 & 0.0257695(31) & 33 & 17 & 12.8 & 3840 & 1 & 20 & 7.6 \\ %
A3 & $1920 \times 960 $ & 1.78 & 0.0176104(31) & 34 & 17 & 12.8 & 7680 & 1 & 10 & 7.6 \\ %
A4 & $2880 \times 1440$ & 1.85 & 0.0112608(29) & 32 & 16 & 12.8 & 30720 & 1 & 2.5 & 7.68 \\ %
\midrule
B1 & $5760 \times 1440$ & 1.85 & 0.0112607(73) & 65 & 16 & 12.8 & 7680 & 1 & 1.5 & 1.152 \\ %
B2 & $2880 \times 2880$ & 1.85 & 0.0112462(72) & 32 & 32 & 12.8 & 7680 & 1 & 1.5 & 1.152 \\ %
\bottomrule
\end{tabular}
\caption{\label{t:ens} Details for the ensembles of field configurations generated for this work. The dimensionful scale $m_{\star}\simeq m$ is defined in Eq.~\eqref{e:m}. Ensembles A1-A4 enable the continuum limit at approximately fixed physical volume, while B1 and B2 are used to estimate finite size effects. Each ensemble consists of $N_{\rm rep}$ independent identical replica, each of which is thermalized for $N_{\rm th}$ cluster updates before $N_{\rm bin}$ measurements are taken by averaging over $B$ subsequent updates. This results in a total number of measurement cluster updates $N_{\rm c}$ for ensemble. Only two digits are given for $m_\star L$ and $m_\star T$.}
\end{table}

The numerical simulations are performed with the single-cluster algorithm described in App.~\ref{a:alg}. Details concerning the generated ensembles of field configurations are summarized in Tab.~\ref{t:ens}. The ensembles A1, A2, A3, A4 have different values of $\beta$ and therefore different values of the lattice spacing, but similar $m_\star L \ge 29$ and $m_\star T \ge 14$. The ensembles B1 and B2 have been generated with the same lattice spacing as A4 but with doubled spatial and temporal extent, respectively. While the ensembles A1, A2, A3, and A4 are used to perform a continuum extrapolation, B1 and B2 enable estimates of the residual finite-$L$ and finite-$T$ effects, as explained
in Sec.~\ref{s:res}.

\begin{figure}
\includegraphics[width=\textwidth]{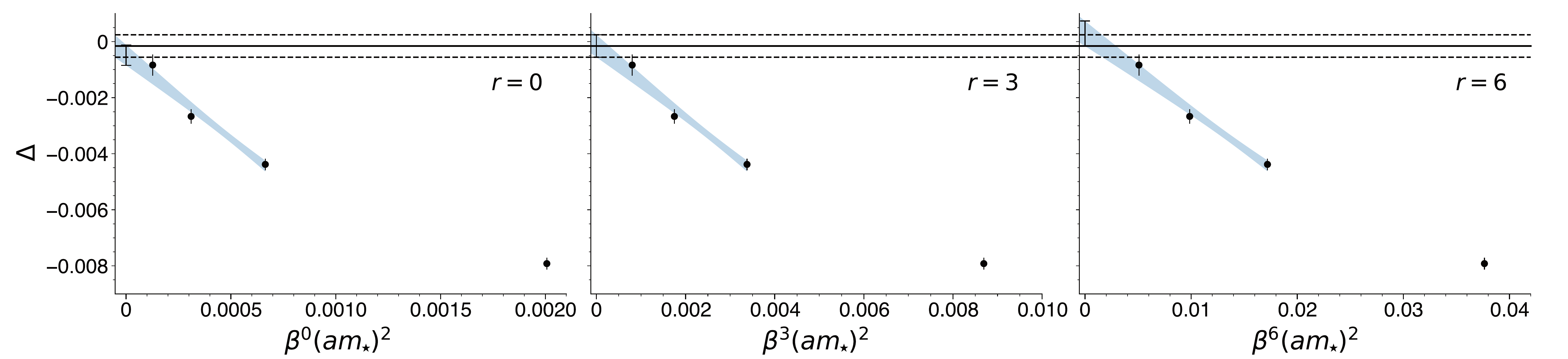}
\caption{\label{f:ccont} Illustration of the continuum limit procedure for $tC(t)|_{m_{\star}t=0.5}$. Three different extrapolation forms are employed using Eq.~\eqref{eq:lat:cont-pheno} with $r=0,3,6$. Shown is the difference between the numerical estimate and the exact result divided by the exact result, denoted $\Delta$. The horizontal band indicates our estimate for the continuum extrapolated $\Delta$, given by the intercept from the fit with $r=3$. The statistical error on this estimate is similar to the variation across the three values of $r$, which are taken as an estimate of the systematic error. This validates the use of the `phenomenological' extrapolation forms in Eq.~\eqref{eq:lat:cont-pheno} for $\rho_{\epsilon}(E)$ where the statistical errors are larger. }
\end{figure}

The standard discretization of the two-dimensional O(3) $\sigma$-model employed here is known to approach the continuum limit rather slowly. Ref.~\cite{Hasenbusch:2001ht} observed that lattice artifacts behave like $O(a)$ over a large range of lattice spacings, in apparent contradiction with Symanzik's effective theory, which predicts an asymptotic $O(a^2)$ behavior up to logarithms. The puzzle was solved in Refs.~\cite{Balog:2009yj, Balog:2009np}: the asymptotic behavior is correctly described by Symanzik's effective theory, but the logarithmic corrections turn out to be large and must be included in fitting formulae used to extrapolate to the continuum limit. On-shell quantities, such as the mass gap or energy levels in finite volume, have the asymptotic expansion
\begin{gather}
Q(a) = Q(0) + C a^2 \beta^3 \left[ 1 + \sum_{k=1}^\infty c_k \beta^{-k} \right] + O(a^4)
\, ,
\label{eq:lat:cont-onshell}
\end{gather}
where $c_1$ and $c_2$ are universal and calculable in terms of two-loop and three-loop integrals respectively ($c_1=-1.1386\dots$ is analytically known and $c_2=-0.4881$ is estimated numerically in Ref.~\cite{Balog:2009np}), while the other constants are non-perturbative. The leading logarithm $\beta^3$ is generated by the dimension-four operator with largest one-loop anomalous dimension appearing in the Symanzik expansion of the action. The current two-point function, and consequently its spectral density, get extra contributions from dimension-three operators appearing in the Symanzik expansion of the current. These one-loop anomalous dimensions are unknown, and their calculation is well beyond the scope of this work. For the continuum extrapolation of the smeared spectral densities we use a fit function of the type
\begin{gather}
Q(a) = Q(0) + C a^2 \beta^r
\, ,
\label{eq:lat:cont-pheno}
\end{gather}
where the exponent $r$ is fixed to $0,3,6$, and $Q(0)$ and $C$ are fit parameters. We take $Q(0)$ at $r=3$ as our continuum extrapolation and the spread generated by the three values of $r$ as an estimate of the systematic error. One may argue that, since a $\beta^3$ term exists in the Symanzik expansion, the correct asymptotic formula should have $r \ge 3$. However it is conceivable that the coefficients of the various logarithms conspire in such a way that an effective power $r<3$ is generated in some intermediate regime. We therefore choose to include also $r=0$ in our analysis. As is evident in Fig~\ref{f:ccont}, the variation of $r$ across $0$, $3$, and $6$ has little effect on the continuum extrapolation of $tC(t)$ with $t=0.5m_{\star}^{-1}$ using our three finest lattice spacings. These conclusions are supported by additional extrapolations at several values of $t$ in the range $t \in [0.5m_{\star}, 4m_{\star}]$. This strategy is thus adopted in Sec.~\ref{s:res} for continuum extrapolations of $\rho_{\epsilon}(E)$, where the fourth lattice spacing is included for stability in the presence of larger statistical errors on $\rho_{\epsilon}(E)$.

%%%%%%%%%%%%%%%%
%%           FIGURE 1          %%
%%%%%%%%%%%%%%%%
%%%%%%%%%%%%%%%%
%%%%%%%%%%%%%%%%
%%%%%%%%%%%%%%%%

\section{Numerical Results}\label{s:res}

After detailing estimates of the systematic errors due to the reconstruction and finite volume, this section presents the numerical verification of the spectral reconstruction procedure using two different tests. The first test (discussed in Sec.~\ref{s:fixed}) compares $\rho_{\epsilon}(E)$ with exact results at fixed smearing width $\epsilon$ for each of the four smearing kernels in Eq.~\eqref{e:kers}. As anticipated, $\rho_{\epsilon}(E)$ is more difficult to determine with increasing $E$ and decreasing $\epsilon$. The second test, which is detailed in Sec.~\ref{s:scale}, uses $\rho_{\epsilon}(E)$ at finite $\epsilon$ to extrapolate to the $\epsilon\rightarrow 0$ limit and obtain the unsmeared spectral density $\rho(E)$. The extrapolation procedure (as applied here) requires smearing widths that are small compared to the scale at which the unsmeared spectral density varies, and is therefore less effective in the elastic region where $\rho(E)$ increases rapidly with the onset of two-particle phase space. Since $\rho(E)$ in the elastic region is accessible to the finite-volume formalism, the focus is instead on energies in the inelastic region $E>4m$. Due to the absence of any sharp `resonance peaks', $\rho(E)$ varies increasing slowly with increasing $E$, suggesting that the smearing width should be scaled $\epsilon \propto (E-2m)$, i.e. with the distance to the rapid variation from the two-particle threshold. Apart from this scaling of the smearing width, the analysis for these two tests proceeds identically.

All statistical errors on the results presented here are estimated using the bootstrap procedure with $N_{\rm boot}=800$  samples, which are generated independently on each ensemble from the $N_{\rm rep} \times N_{\rm bin}$ measurements listed in Tab.~\ref{t:ens}. For all reconstructions the lower bound of the integration range $E_0$ defined in Eq.~\eqref{eq:AandB} is set using the scale $m_{\star}$ via $aE_0 = 2\,am_{\star}$ to (approximately) coincide with the two-particle threshold. This choice minimizes the range in energy over which the functional $A[g]$ forces the reconstructed and target kernels to be similar. The values of $am_{\star}$ given in Tab.~\ref{t:ens} are also used to fix the dimensionful parameters $\epsilon$ and $E$ at different lattice spacings. Properly including the statistical error on $am_{\star}$ requires the determination of the coefficients $g_{\tau}^{\lambda}$ on each bootstrap sample, which significantly increases the computational cost. After confirming that this has no observable effect on a selection of sample reconstructions, the statistical error on the values of $am_{\star}$ in Tab.~\ref{t:ens} is subsequently ignored.

\subsection{Fixed smearing width}\label{s:fixed}

\begin{figure}
\includegraphics[width=\textwidth]{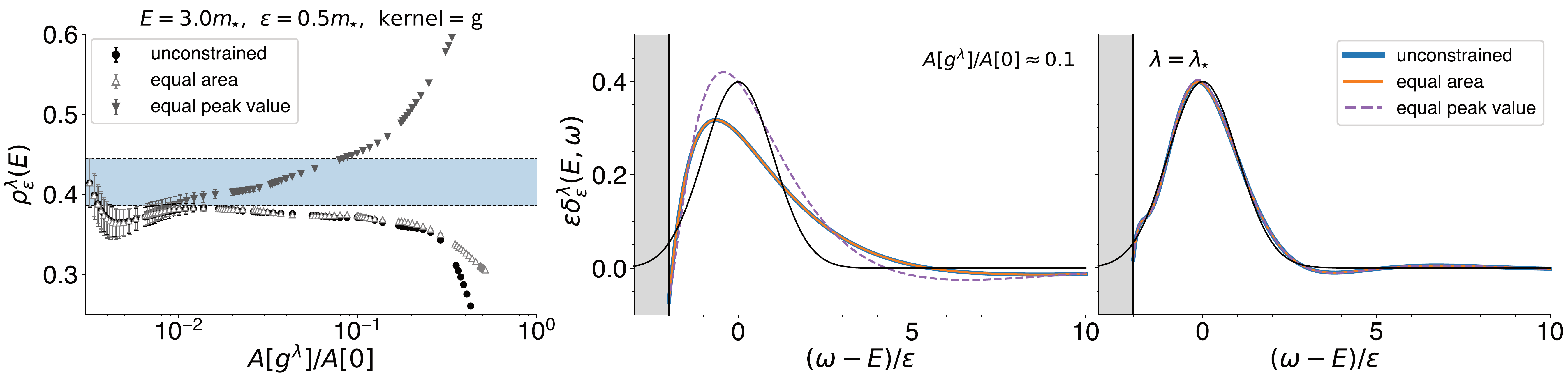}
\caption{\label{f:const} For the A1 ensemble detailed in Tab.~\ref{t:ens}, an illustration of the compromise between statistical and systematic errors which results from choosing $\lambda=\lambda_\star$. {\bf Left}:
the estimator $\rho^{\lambda}_{\epsilon}(E)$ for the smeared spectral function
at different values of $\lambda$, plotted against the `resolution' functional $A[g^{\lambda}]/A[0]$ defined in Eq.~\eqref{eq:AandB}. The horizontal band indicates
the result at $\lambda_\star$ (where $A[g^\star] = A[0]B[g^\star]$) without any constraints on the
determination of the optimal coefficients. This choice is consistent
with the one obtained by imposing the equal area constraint (denoted `equal area') as well as with the one demanding that reconstructed smearing kernel
	take the correct value at the peak, denoted `equal peak value'. {\bf Center}: the reconstructed smearing kernel $\delta_{\epsilon}^{\lambda}(E,\omega)$ for the different constraints together with the target kernel $\delta_\epsilon^{\sf g}(E-\omega)$ shown as a solid line. The value of $\lambda$ for each of the three reconstructions results in $A[g^{\lambda}]/A[0] \approx 0.1$ and the vertical shaded band indicates energies $\omega < E_0$ where the reconstructed kernels are left unconstrained. {\bf Right}: same as the center panel but with $\lambda = \lambda_{\star}$.}
\end{figure}

\begin{figure}
\includegraphics[width=\textwidth]{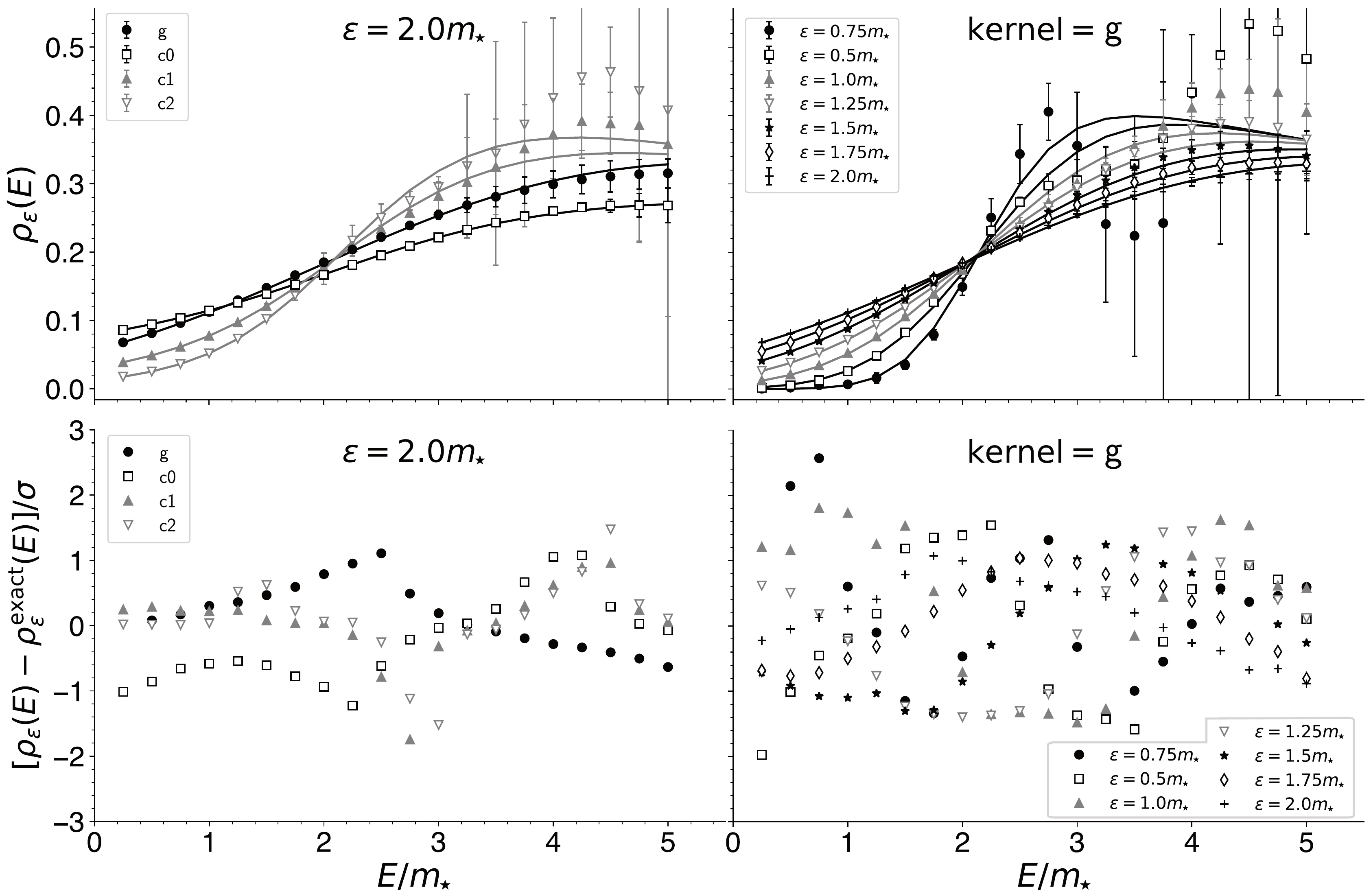}
\caption{\label{f:efix} Summary of the continuum extrapolated values with statistical and systematic errors combined in quadrature as described in the text. {\bf Top left}: all four smearing kernels are plotted for a large smearing radius $\epsilon = 2m_{\star}$ together with the exact results including 2-, 4-, and 6-particle contributions shown as solid lines. {\bf Bottom left}: the `pull' between the numerical and exact results for the same data. The variation of the points indicates that the numerical and exact results are consistent at the 2-$\sigma$ level. {\bf Top and bottom right:} same type of plots but for a variety of smearing widths with the Gaussian smearing kernel.}
\end{figure}

Before presenting the main results of this section in Fig.~\ref{f:efix}, several systematic errors must be estimated. Consider first systematic errors due to the reconstruction procedure. Due to the finite number of input values $C(a\tau)$, the reconstructed smearing kernel will never be exactly equal to the desired one. This source of systematic error is quantified by the functional $A[g^\lambda]/A[0]$ defined in Eq.~\eqref{eq:AandB}. While this measure does not take into account the role of the unsmeared spectral density $\rho(E)$ on the systematic error of $\rho_{\epsilon}(E)$, this error certainly vanishes in the $A[g^\lambda]\rightarrow 0$ limit. The value of $A[g^\lambda]$ is therefore a useful diagnostic for determining the onset of the statistics-limited regime: if $\lambda$ is lowered such that $A[g^\lambda]$ changes significantly and no significant change is observed in $\rho^\lambda_{\epsilon}(E)$, then this source of systematic error is likely smaller than the statistical error. This type of `plateau' analysis is well-known to lattice field theory practitioners and is exemplified in Fig.~\ref{f:const}. As discussed in Sec.~\ref{s:exact}, we follow here the recipe of Ref.~\cite{Hansen:2019idp} and quote as the best estimate for $\rho_\epsilon(E)$ the result from the unconstrained reconstruction obtained at $\lambda=\lambda_\star$ defined in Eq.~\eqref{eq:Wmax}.

%%%%%%%%%%%%%%%%
%%           FIGURE 2          %%
%%%%%%%%%%%%%%%%
%%%%%%%%%%%%%%%%
%%%%%%%%%%%%%%%%
%%%%%%%%%%%%%%%%

Another probe of the systematic error due to the reconstruction is the comparison of $\rho_{\epsilon}(E)$ determined from coefficients subject to various constraints on the minimization, implemented as explained in App.~\ref{a:hlt}. Fig.~\ref{f:const} shows results for the unconstrained reconstruction, the constraint that the reconstructed smearing kernel has a signed area equal to the desired one, defined in Eq.~\eqref{eq:cost1}, and the constraint that the reconstructed kernel exactly coincides with the desired one at the peak $\omega = E$, defined in Eq.~\eqref{eq:cost2}. For large $\lambda$, these different reconstructions differ significantly\footnote{The equal area constraint is apparently somewhat weaker than demanding $\delta_{\epsilon}^{\lambda}(E,\omega) = \delta_{\epsilon}^{\sf x}(E-\omega)$ at $\omega = E$, as evidenced by the plots of $\delta_{\epsilon}^{\lambda}(E,\omega)$ in the center and left panels of Fig.~\ref{f:const}.} but coincide well within the statistical error at $\lambda_\star$, lending additional confidence that $\lambda_\star$ is indeed in the statistics-dominated regime. Put more precisely, reconstructions subject to different constraints result in different reconstructed smearing kernels. Their consistency when compared at similar values of $A[g^{\lambda}]/A[0]$ demonstrates that these differences are insignificant and suggests that deviations from the exact kernel are as well. Overall, the approach employed for choosing $\lambda_{\star}$ discussed in Eq.~\eqref{eq:Wmax} to balance statistical and systematic errors is somewhat conservative. Fig~\ref{f:const} suggests the less restrictive alternative strategy of merely demanding consistency between the three different reconstructions. This could potentially result in smaller statistical errors, but requires further investigation beyond the scope of this work.

%%%%%%%%%%%%%%%%
%%           FIGURE 3          %%
%%%%%%%%%%%%%%%%
\begin{figure}
\includegraphics[width=0.8\textwidth]{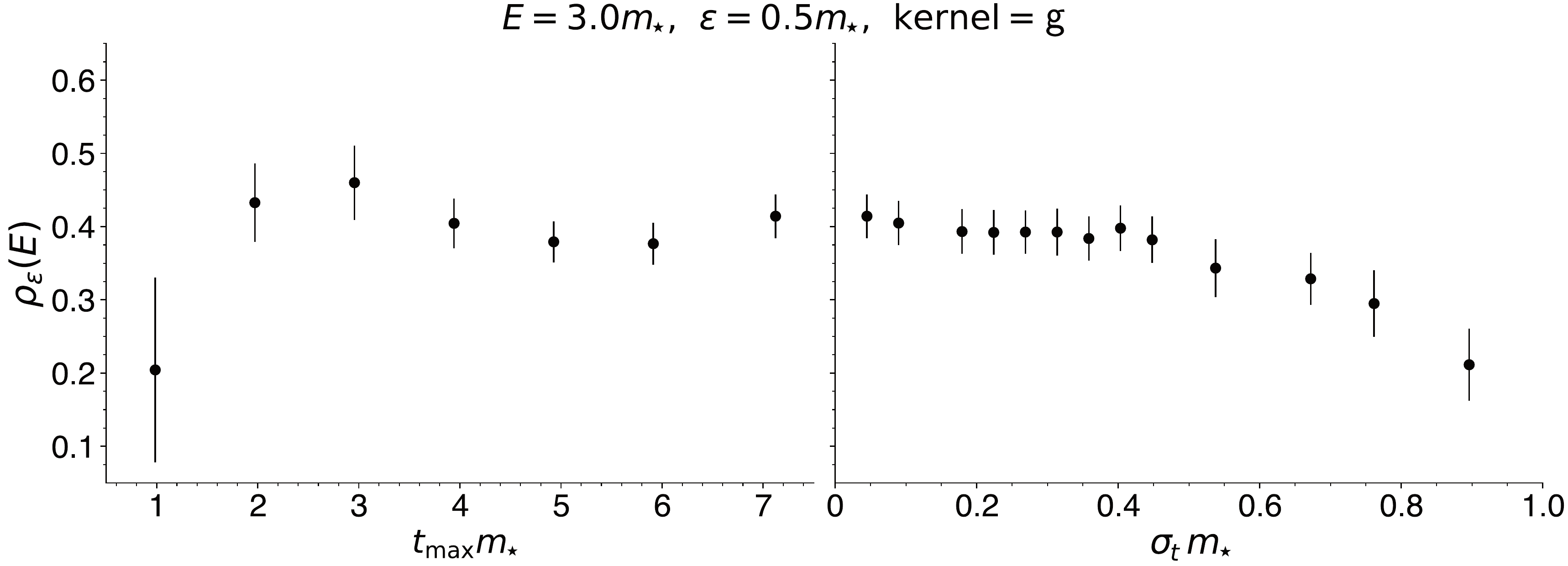}
\caption{\label{f:rtest} {\bf Left}: test of the sensitivity of the reconstruction procedure to the range of time slices $[\tau_{\rm min}, \tau_{\rm max}]$ by varying $\tau_{\sf max}$, plotted using $t_{\rm max} m_{\star} = \tau_{\sf max} (a m_{\star})$ with fixed $\tau_{\rm min} = 1$. {\bf Right}: test of the sensitivity to the spacing between time slices included in the
reconstruction $\sigma_{t}$. Both tests are performed on the A1 ensemble detailed in Tab.~\ref{t:ens}. Evidently the reconstruction procedure is relatively stable under the variation of $\sigma_{t}$ and $\tau_{\rm max}$.
}
\end{figure}
%%%%%%%%%%%%%%%%
%%%%%%%%%%%%%%%%
%%%%%%%%%%%%%%%%

The input correlator time slices $\{C(a\tau)\}$ range over $[\tau_{\rm min}, \tau_{\rm max}]$ with $\tau_{\rm min} = 1$ and $\tau_{\rm max} = 160$ used as the best estimate of $\rho_{\epsilon}(E)$ everywhere. They are furthermore correlated and suffer from an exponential degradation of the signal-to-noise ratio with increasing $\tau$ with a rate phenomenologically similar to $m$.
The sensitivity to the input correlator data is probed in Fig.~\ref{f:rtest} 
which demonstrates that, for a sample reconstruction, $\rho_{\epsilon}(E)$ is relatively insensitive to $\tau_{\rm max}$ and to a `thinning' of the input data where the time slices are separated by $\sigma_{t} = a(\tau_{n+1} - \tau_n)$. These observations can be plausibly explained by the signal-to-noise degradation and correlation of the input data. The role of $B[g]$ is to penalize coefficients $g_{\tau}^{\lambda}$ which would result in a large statistical error. This penalty naturally disfavors input data at large $\tau$, effectively `turning off' these time slices and resulting in an insensitivity to $\tau_{\rm max}$. Similarly, the correlation between the input time slices may be responsible for the robustness to changing $\sigma_{t}$.

\bigskip

\begin{figure*}
\includegraphics[width=\textwidth]{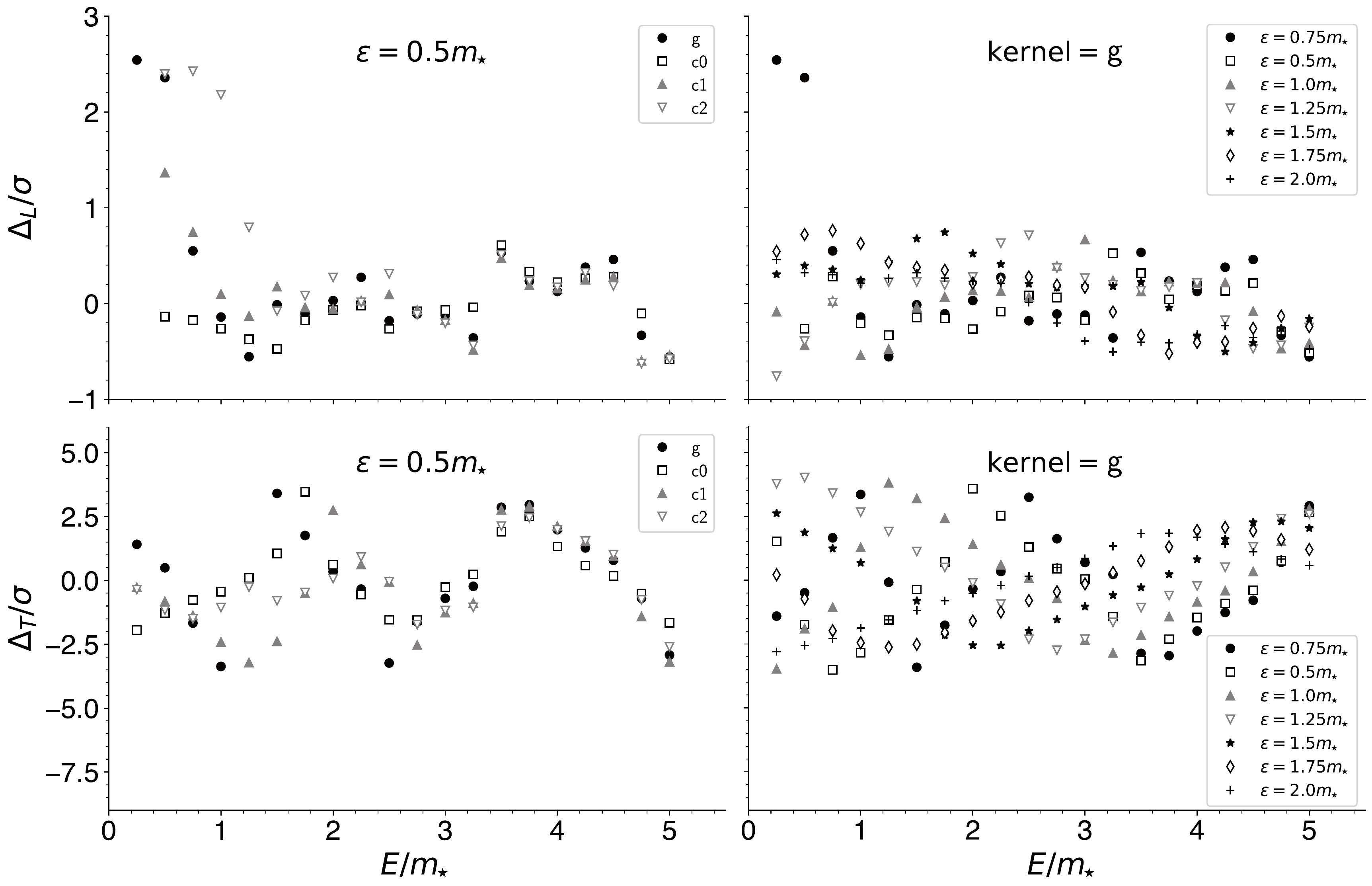}
\caption{\label{f:ltest}The differences $\Delta_{\rm L}$ and $\Delta_{\rm T}$ from Eq.~\eqref{e:delta}. The vertical axis is normalized by the error on the estimated difference and can be interpreted as the statistical significance of each of the finite size effects. The sum of the mean values $|\Delta_{\rm L}| + |\Delta_{\rm T}|$ is taken as an estimate of the systematic error due to finite size effects independently for each $E$, $\epsilon$, and smearing kernel.}
\end{figure*}

The preceding discussion demonstrates that the reconstruction procedure on a particular ensemble of field configurations results in a statistics-dominated estimator for $\rho_{\epsilon}(E)$. No further systematic error due to the 
reconstruction is assigned. We turn now to systematic errors associated with the finite extent of the lattice. Along the (approximate) line of constant physics defined by the ensembles A1-A4, these errors are crudely estimated by considering separately the differences between B1 and A4, and B2 and A4. These differences
\begin{equation}\label{e:delta}
\Delta_{\rm L}(\epsilon,E) = \rho_{T,L,\epsilon}(E) - \rho_{T,2L,\epsilon}(E)
\qquad {\rm and} \qquad
\Delta_{\rm T}(\epsilon,E) = \rho_{T,L,\epsilon}(E) - \rho_{2T,L,\epsilon}(E) \,,
\end{equation}
are taken as estimators for the systematic errors due to the finite $L$ and $T$ and are added (in absolute value) independently for each $\epsilon$, $E$, and smearing kernel. A selection of these deviations are shown in Fig.~\ref{f:ltest}, illustrating that they are at most marginally significant. Nonetheless, this systematic error is subsequently taken into account and represents the largest of the systematic error estimates. Although Eq.~\eqref{eq:rhoTL-Tinf} ensures that finite-$T$ effects are `reliably' ${\rm O}({\rm e}^{-mT})$, we conservatively account for both finite $L$ and $T$ with additional systematic errors as discussed above. An increase of statistics on the B1 and B2 ensembles would more accurately estimate these systematic errors and provide a more stringent test of the error estimates due to the reconstruction, but is left for future work.

%%%%%%%%%%%%%%%%
%%           FIGURE 4          %%
%%%%%%%%%%%%%%%%
%%%%%%%%%%%%%%%%
%%%%%%%%%%%%%%%%
%%%%%%%%%%%%%%%%

\begin{figure}
\includegraphics[width=\textwidth]{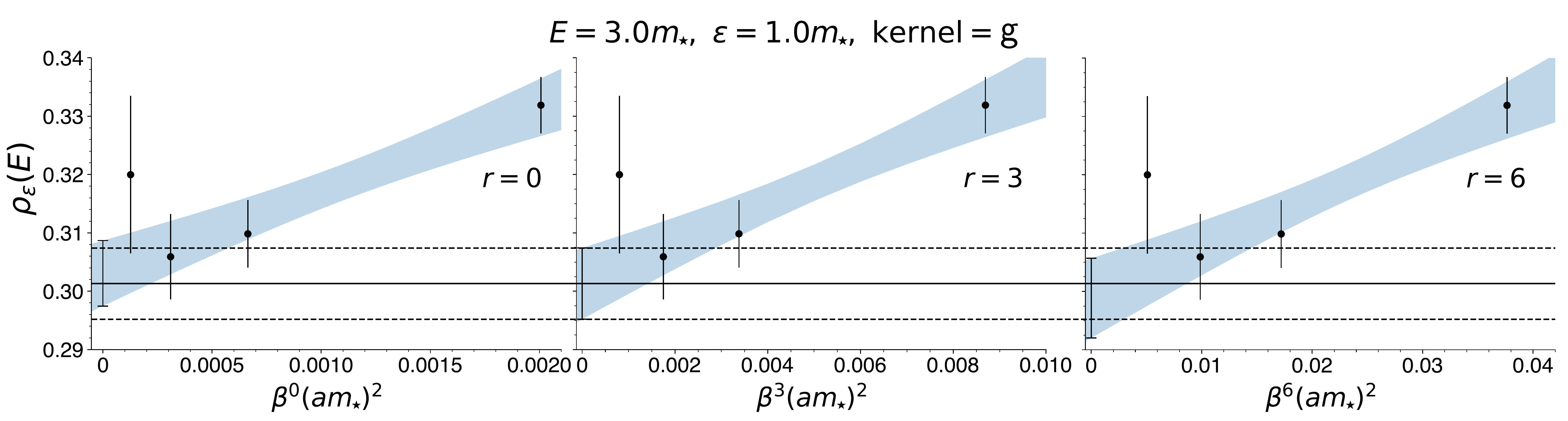}
\caption{\label{f:ctest} Illustration of the continuum extrapolation
procedure from Eq.~\eqref{eq:lat:cont-pheno} with $r = 0$, $3$, and $6$ at $E=3m_{\star}$, $\epsilon=m_{\star}$, and the Gaussian smearing kernel. The three panels show the $r=0$ ansatz (left), the $r=3$ (center), and the $r=6$ (right). The points at $a=0$ are the continuum extrapolated values and the horizontal band is the extrapolated value from $\beta^3a^2$, which is taken as the best estimate. As in Fig.~\ref{f:ccont}, there is evidently little variation across the different extrapolation forms. }
\end{figure}

In addition to finite size effects, systematic errors due to the lattice spacing must be estimated as outlined in Sec.~\ref{s:lat}. The continuum limit is taken with four lattice spacings using the ensembles A1-A4, but (as detailed in Sec.~\ref{s:lat}) the asymptotic $a^2$ behavior is affected by large logarithmic corrections and the `phenomenological' extrapolation form of Eq.~\eqref{eq:lat:cont-pheno} is employed with $r=0$, $3$, and $6$. The value at $r=3$ is taken as the best estimate and the largest deviation between any two as an estimate of the systematic error.
A selection of the extrapolations for the three different values of $r$ is shown in Fig.~\ref{f:ctest}. Evidently, the difference in the extrapolated value varies little across these three extrapolation forms.

%%%%%%%%%%%%%%%%
%%           FIGURE 5          %%
%%%%%%%%%%%%%%%%
%%%%%%%%%%%%%%%%
%%%%%%%%%%%%%%%%
%%%%%%%%%%%%%%%%

After estimating systematic errors due to the reconstruction, finite lattice size, and finite lattice spacing, the results for $\rho_{\epsilon}(E)$ are at last confronted with the exact values. While no further systematic error is assigned to the reconstruction procedure, estimates of the remaining three sources of systematic error (finite $L$, finite $T$, continuum limit) are added naively. The total systematic error is then combined in quadrature with the statistical error on the continuum limit estimator. The finite-$L$ and $T$ errors are typically the dominant source of systematic error, and are similar in magnitude to the statistical error.

%%%%%%%%%%%%%%%%
%%           FIGURE 6          %%
%%%%%%%%%%%%%%%%
%%%%%%%%%%%%%%%%
%%%%%%%%%%%%%%%%
%%%%%%%%%%%%%%%%
A selection of these continuum extrapolated values are shown in Fig.~\ref{f:efix}. In the left two panels all four kernels are compared at a large fixed $\epsilon = 2m_{\star}$ to accentuate the difference between them. However, all of these results are consistent with the exact values (including 2-, 4-, and 6-particle contributions) within the combined statistical and systematic errors. The right two panels show various $\epsilon$ for the Gaussian kernel and are similarly consistent with the exact numbers. The increasing difficulty of the reconstruction problem with increasing $E$ and decreasing $\epsilon$ is apparent.

\subsection{Extrapolation to zero smearing width}\label{s:scale}

The continuum extrapolated results for $\rho_{\epsilon}(E)$ displayed in Fig.~\ref{f:efix} demonstrate that the spectral reconstruction procedure yields quantitatively accurate results within the quoted statistical and systematic errors. However, the difficulty in reconstructing $\rho_{\epsilon}(E)$ at large $E$ and small $\epsilon$ naively suggests that the approach is of little use in the inelastic region, precisely where the finite-volume formalism is not yet developed. However, the increasing smoothness of $\rho(E)$ with increasing $E$ can be exploited by scaling $\epsilon \propto (E-2m_{\star})$ in order to probe larger energies. The scaled values of $\rho_{\epsilon}(E)$ are used to extrapolate $\epsilon \rightarrow 0$ and determine the unsmeared spectral density $\rho(E)$ using the small-$\epsilon$ expansion discussed in Sec.~\ref{s:exact}. The rapid variation of $\rho(E)$ in the elastic region inhibits the application of this approach there, so we focus on $E\ge4m_{\star}$\footnote{The opening of subsequent $2n$-particle thresholds introduces non-analyticities in $\rho(E)$, but for the range of energies considered here $\rho(E)$ is dominated by fewer-particle contributions and therefore approximately smooth.}.  

%%%%%%%%%%%%%%%%
%%           FIGURE 7          %%
%%%%%%%%%%%%%%%%
\begin{figure}
\includegraphics[width=\textwidth]{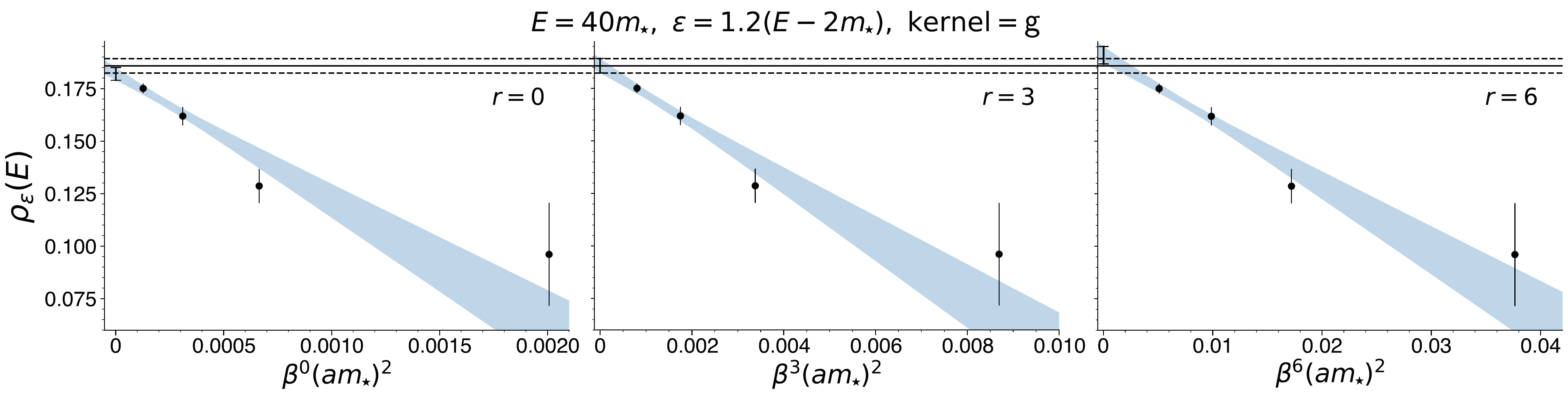}
\caption{\label{f:cont2}The same as Fig.~\ref{f:ctest} for $E = 40m_{\star}$, $\epsilon = 1.2(E-2m_{\star})$, and the Gaussian smearing kernel.}
\end{figure}
%%%%%%%%%%%%%%%%
%%%%%%%%%%%%%%%%
%%%%%%%%%%%%%%%%

A natural concern is the growth of cutoff effects with increasing $E$ and $\epsilon$, since $\rho_{\epsilon}(E)$ has increasingly significant contributions from more energetic states. Fig.~\ref{f:cont2} shows the continuum extrapolations of $\rho_{\epsilon}(E)$ at the largest energy considered here, which is $E=40m_{\star}$, and $\epsilon = 1.2(E-2m_{\star})$, which is the largest $\epsilon$ used in the $\epsilon \rightarrow 0$ extrapolation. Although the continuum limit is somewhat steeper than those in Fig.~\ref{f:ctest}, there is still little variation across the different values of $r$. However, it appears that the statistical error on the estimator $\rho_{\epsilon}^{\lambda}(E)$ decreases more rapidly with $am_{\star}$ in Fig.~\ref{f:cont2} than in Fig.~\ref{f:ctest}. This phenomenon requires further study, but could be influenced by our choice of basis vectors, namely the inclusion of all $\tau \in [1,160]$ at each lattice spacing. The edge of the Brillouin zone may also play a role since $1/(am_{\star}) \simeq E/m_{\star}$ when $E \simeq 20m_{\star}$ for the A1 ensemble and $E \simeq 90m_{\star}$ for the A4. Although not displayed here, the continuum-extrapolated values for the energies and smearing widths used in this section are consistent with the exact results at a level similar to Fig.~\ref{f:efix}.

The small-$\epsilon$ behavior of $\rho_{\epsilon}(E)$ depends on the smearing kernel\footnote{For a fair comparison of the effectiveness of the four kernels in Eq.~\eqref{e:kers}, $\epsilon$ for the {\sf c2} is enlarged by the rescaling $\epsilon \rightarrow \sqrt{3}\epsilon$ so that $w_2^{\sf c2} = 1$ in Tab.~\ref{t:wcoeffs}. This rescaling is not employed for the {\sf c2} data in Sec.~\ref{s:fixed}}. As discussed in Sec.~\ref{s:exact}, for the four smearing kernels considered here, this dependence is encoded in the coefficients $w_{k}^{\sf x}$, which are known analytically and given in Tab.~\ref{t:wcoeffs}. The coefficients $a_k(E)$ in Eq.~\eqref{e:expand} however depend on $\rho(E)$ only and are therefore identical across the different smearing kernels. This means that an extrapolation ansatz which includes terms up to $O(\epsilon^n)$ contains $n+1$ fit parameters for each of the $a_k$, which are constrained by data from all the smearing kernels. In this work we target $a_{0}(E) = \rho(E)$ only: no additional information from the other $a_k(E)$ is used to further constrain the desired spectral density.

%%%%%%%%%%%%%%%%
%%           FIGURE 8          %%
%%%%%%%%%%%%%%%%
\begin{figure}
\includegraphics[width=0.95\textwidth]{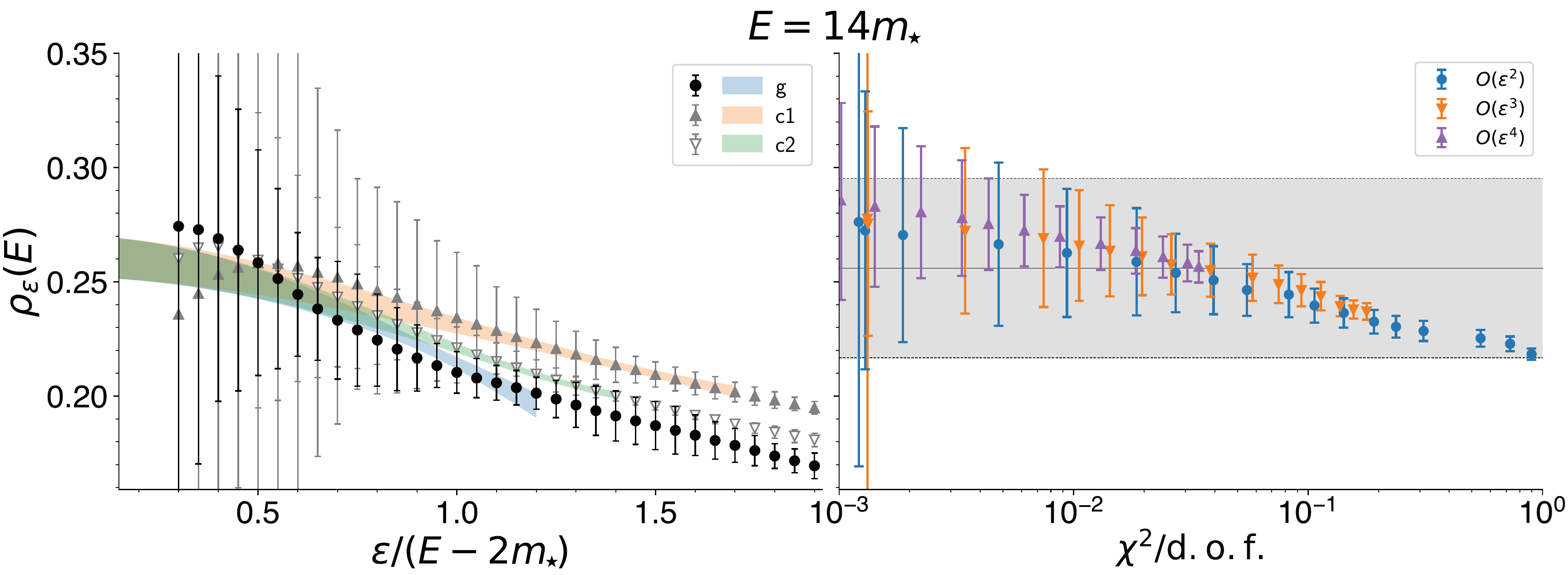}
\caption{\label{f:extrap} {\bf Left}: a sample constrained extrapolation of
the {\sf g}, {\sf c1}, and {\sf c2} kernels at $E=14m_{\star}$ with the (constrained)
extrapolation form including terms up to $O(\epsilon^{4})$. {\bf Right}: summary of the extrapolated values for different extrapolations forms and different extrapolation ranges $[\epsilon_{\rm min}, \epsilon_{\rm max}]$ with $\epsilon_{\rm min} = 0.3(E-2m_{\star})$ fixed and varying $\epsilon_{\rm max}$. For each $\epsilon_{\rm max}$, the extrapolated result is plotted with the correlated $\chi^2/{\rm d.o.f.}$ as the horizontal coordinate. The horizontal band shows the final estimate for the extrapolated values, with statistical and systematic errors combined in quadrature. The systematic errors are estimated as discussed in the text.}
\end{figure}
%%%%%%%%%%%%%%%%
%%%%%%%%%%%%%%%%
%%%%%%%%%%%%%%%%

The clear sources of systematic error in this approach are the extrapolation form and the extrapolation range $[\epsilon_{\rm min}, \epsilon_{\rm max}]$. Due to the difficulty in reconstructing small $\epsilon$, the statistical errors increase with decreasing $\epsilon$ so that in practice varying $\epsilon_{\rm min}$ has little effect on the extrapolated value. We therefore fix $\epsilon_{\rm min} = 0.3(E-2m_{\star})$ for all smearing kernels and extrapolations. The effect of varying $\epsilon_{\rm max}$ must be monitored more closely, however. In order to fairly include data from different smearing kernels, $\epsilon_{\rm max}$ for each kernel is re-scaled $\epsilon_{\rm max} \rightarrow \epsilon_{\rm max}/\alpha_{\sf x}$ with $\int_{-\alpha_{\sf x}}^{\alpha_{\sf x}}{\rm d}x \, \delta^{\sf x}_{\epsilon=1}(x) = {\rm erf}(2^{-1/2}) \approx 0.68269$ chosen to approximately coincide with the Gaussian. This treats the data from different smearing kernels on equal footing: all smearing widths are included up to those with the same amount of `leakage' down to the two-particle threshold, where $\rho(E)$ varies rapidly. For the {\sf c0} kernel the large value $\alpha_{\sf c0} \approx 1.84$ together with the leading $O(\epsilon)$ behavior effectively renders it useless in the extrapolations. For the Gaussian kernel $\alpha_{\sf g} = 1$ by definition, while $\alpha_{\sf c1} = 0.7$ and $\alpha_{\sf c2} = 0.855$ increase the fit range for these two kernels relative to the Gaussian. The extrapolations are therefore performed with the three kernels {\sf g}, {\sf c1}, and {\sf c2}. An example of such an extrapolation is shown in the left panel of Fig.~\ref{f:extrap}.

\begin{figure}
\includegraphics[width=0.6\textwidth]{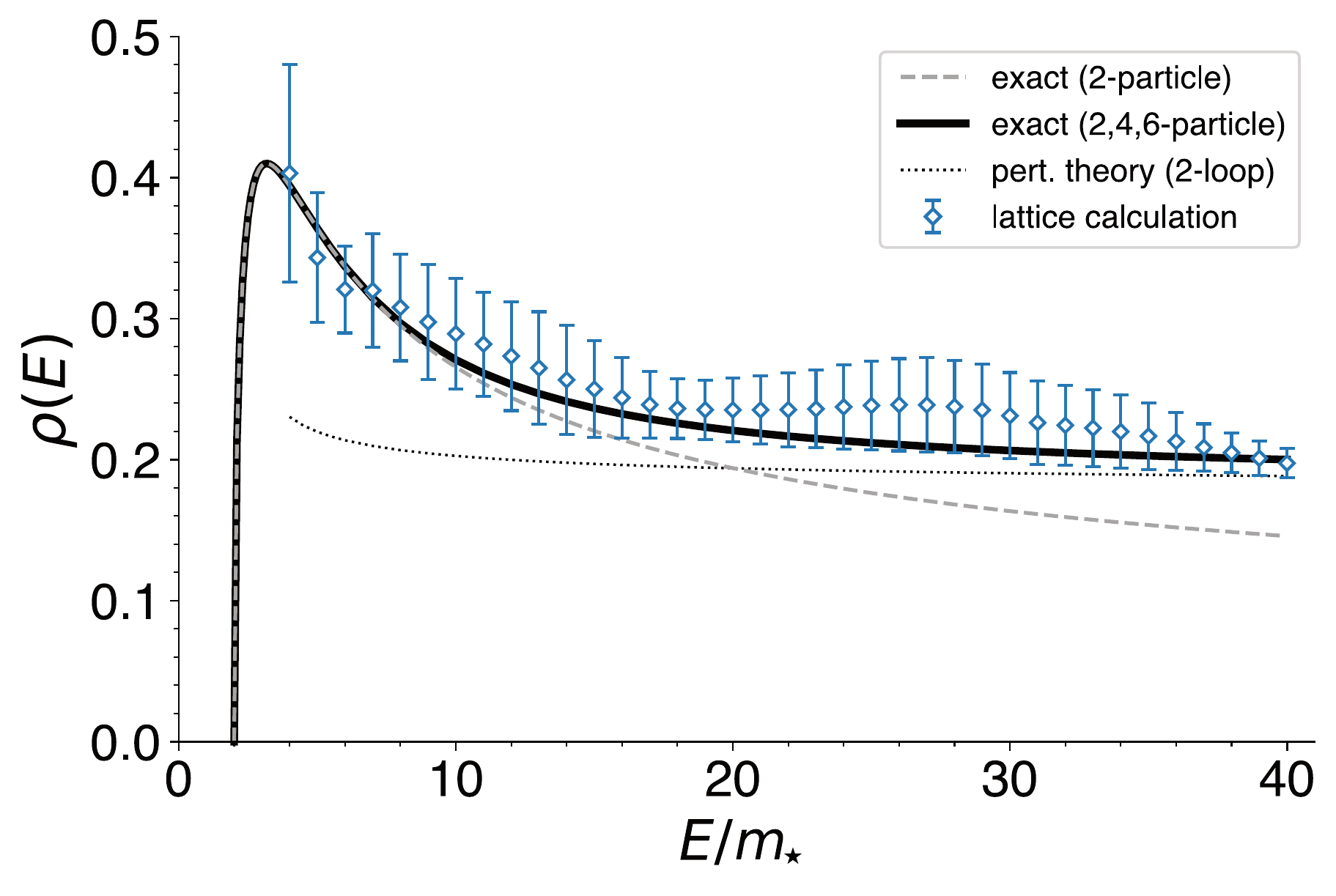}
\caption{\label{f:inel}A summary of the continuum, $\epsilon\rightarrow 0$ extrapolated
	results for $\rho(E)$, together with the exact two-particle contribution (light dashed line), the two-, four-, and six-particle contributions (dark solid line), and the 2-loop perturbative result (dark dotted line). Statistical and systematic errors due to the finite volume, continuum limit, and $\epsilon \rightarrow 0$ extrapolation are combined in quadrature as described in the text.}
\end{figure}

The systematic error estimate due to the $\epsilon \rightarrow 0$ extrapolation proceeds by employing extrapolation forms up to and including $O(\epsilon^p)$ with $p = 2$, $3$, and $4$ and varying $\epsilon_{\rm max}$ keeping below $\epsilon_{\rm max} < 1.3(E-2m_{\star})$. For each value of $p$, the largest $\epsilon_{\rm max}$ resulting in a (correlated) $\chi^2/{\rm d.o.f.} < 1$ is identified, with the $p=4$ value taken as the best estimate for $\rho(E)$. The spread of these three values is then added in quadrature as a systematic error. An example plot showing the consistency between different extrapolation forms and extrapolation ranges is shown in the right panel of Fig.~\ref{f:extrap}. This procedure is performed for all energies for which $E/m_{\star}$ takes integral values from $4$ to $40$. A summary of the extrapolated results is shown in Fig.~\ref{f:inel}, which are consistent with the exact results including 2, 4, and 6 particle contributions. Interestingly, the data exhibits some sensitivity to $\rho^{(4)}(E)$ for $E\gtrsim 20m_{\star}$. Four-particle scattering amplitudes are currently beyond the reach of the finite-volume formalism.

%%%%%%%%%%%%%%%%
%%           FIGURE 9          %%
%%%%%%%%%%%%%%%%
%%%%%%%%%%%%%%%%
%%%%%%%%%%%%%%%%
%%%%%%%%%%%%%%%%

\section{Conclusions}\label{s:conc}

The aim of the preceding sections is to verify the procedure of Ref.~\cite{Hansen:2019idp} for numerically computing smeared spectral densities (with an \emph{a priori} specified smearing kernel) from lattice field theory correlation functions. In this regard the two-dimensional O(3) model usefully provides exact results against which the estimates can be checked. These checks, which are presented in Figs.~\ref{f:efix} and~\ref{f:inel}, are satisfied and compare both $\rho_{\epsilon}(E)$ at finite $\epsilon$ and the results from $\epsilon \rightarrow 0$ extrapolations to determine $\rho(E)$ deep into the inelastic region where finite-volume methods have not yet been developed. The highest energy considered here is $E=40m_{\star}$, at which $\rho(E)$ is determined with a relative accuracy of $5\%$ and differs significantly from the exact two-particle contribution $\rho^{(2)}(E)$ given in Eq.~\eqref{e:2part}.

Apart from the `usual' sources of systematic error due to the finite lattice spacing and finite-volume spacetime, we must also consider the imperfect reconstruction of the smearing kernel due to the finite number of input time slices and their associated statistical errors. All sources of systematic error have been estimated and included in Figs.~\ref{f:efix} and~\ref{f:inel} where the statistical and systematic errors are added in quadrature. Generally the errors due to the finite lattice extent are the largest source of systematic uncertainty, and are typically less than or comparable to the statistical errors.

The determination of $\rho_{\epsilon}(E)$ becomes increasingly difficult for smaller smearing widths $\epsilon$ at fixed energy $E$, and increasing $E$ with fixed $\epsilon$. As is evident from the right two panels of Fig.~\ref{f:efix}, it is difficult to achieve precise results outside of the elastic region for $\epsilon \lesssim m/2$ with the Gaussian smearing kernel. Better is to exploit the smoothness of $\rho(E)$ and scale $\epsilon \propto (E-2m)$, so that an equal proportion of the smearing kernel `leaks' down to the two particle threshold at each energy. This enables the determination of $\rho(E)$ in Fig.~\ref{f:inel}, which is the main result of this work.

The analysis performed here is (in principle) directly applicable to the vector-vector correlator in QCD to compute the $R$-ratio, provided a few key points are addressed. First, the presence of narrow resonances in QCD seemingly invalidates our approach to the $\epsilon \rightarrow 0$ extrapolation by scaling $\epsilon \propto (E-2m)$. Nonetheless, the finite smearing width may be used to probe the onset of the perturbative regime~\cite{Poggio:1975af}. Furthermore the spatial extent achieved here of $mL = 30$ is currently beyond the state of the art for simulations at physical quark masses. To this point it is important to note that the density of states is higher in three spatial dimensions compared to one so that smaller values of $mL$ may be acceptable, though still likely larger than the typical lattice QCD volumes. In this respect, the masterfield simulation paradigm~\cite{Francis:2019muy, Giusti:2018cmp} may enable larger volumes in the near future~\cite{Ce:2021akh}.

Apart from this, the difficulty in the reconstruction problem faced here may in fact be similar to QCD. Although the use of $\tau_{\rm max} = 160$ may seem daunting, Fig.~\ref{f:rtest} indicates that reducing the range and density of input correlator time slices likely has little effect. Although the input correlator data for this work is generated using the two-cluster algorithm, it still possesses an exponentially degrading signal-to-noise ratio which decays at a rate empirically similar to $m$, as expected for the vector-vector correlator in QCD. Also, as is typical with scalar field theory, the statistical errors achieved at $C(m_{\pi}^{-1})$ are roughly comparable with state-of-the-art determinations of the vector-vector correlator in QCD.

Finally, the validation of the spectral reconstruction procedure presented here is intended as a stepping stone to other applications. Among them is the determination of exclusive scattering amplitudes by exploiting an asymptotic formalism as in Refs.~\cite{Bulava:2019kbi,Bruno:2020kyl}. There are several additional challenges there compared to this work, including the computation and spectral reconstruction of at least three-point temporal correlation functions with two associated time separations. Nonetheless, this may provide an alternative to the finite-volume formalism for determining few-body scattering amplitudes where applicable, and extend the reach of such computations by enabling (in principle) arbitrary center-of-mass energies.

\acknowledgements{We thank Martin R. Hansen for his participation at an early stage of this work and acknowledge helpful conversations with Rainer Sommer and Mattia Bruno, and with Peter Weisz, who also provided comments on a previous version of this manuscript. We also thank the organizers of the CERN Workshop ``Advances in Lattice Gauge Theory'' where work on this project began. The work of MTH is supported by UK Research and Innovation Future Leader Fellowship MR/T019956/1. The research of AP is funded by the Deutsche Forschungsgemeinschaft (DFG, German Research Foundation) - Projektnummer 417533893/GRK2575 ``Rethinking Quantum Field Theory''. This work was supported by CINECA that granted computing resources on the Marconi supercomputer to the LQCD123 INFN theoretical initiative under the CINECA-INFN agreement.}

\appendix

\section{Analytic expressions for the spectral density}\label{s:app_sd}

Thanks to the integrability of the O(3) model, the matrix elements of the Noether current $j_\mu$ between the vacuum and a generic $n$-particle state (\textit{form factors}) are completely determined by a known set of recursive functional equations, derived by Karowski and Weisz~\cite{Karowski:1978vz}. These functional equations are written in terms of the S-matrix of the system, which has been analytically derived by Zamolodchikov and Zamolodchikov~\cite{Zamolodchikov:1977nu} on the basis of the integrability of the model and some fairly weak assumptions. The $n$-particle contribution to the spectral density of the Noether current can be reconstructed from the form factors. This program has indeed been carried out by Balog and Niedermaier~\cite{Balog:1996ey} for $n=2,4,6$. Interestingly, the original motivation for this calculation was the calculation of the Euclidean two-point function of the O(3) model, and the comparison with lattice simulations and perturbative calculations~\cite{Balog:1996mn}.

For convenience we summarize here the most important formulae taken from~\cite{Balog:1996ey}. A different normalization for the spectral density with respect to Balog and Niedermaier (BN) is employed here
\begin{gather}
\rho(\mu) = \frac{3 \mu}{2} \, \rho_\text{BN}(\mu)
=
\frac{3 \mu}{2} \sum_{n=2,4,\dots} \rho_\text{BN}^{(n)}(\mu)
\ .
\end{gather}
With this normalization, the large-$\mu$ behaviour of the spectral density is given by $\rho(\mu) = 1/(2\pi) + O(g^2(\mu))$.

The form factors are usually written in terms of the rapidities $\theta_{i=1,\dots,n}$. The (spatial) momentum and energy of the $i$-th particle are given by
\begin{gather}
p_i = m \sinh \theta_i
\ , \qquad
E_i = E(p_i) = m \cosh \theta_i
\ .
\end{gather}
The invariant mass of a system of $n$ particles is then
\begin{gather}
M^{(n)}
= \left\{ \left( \sum_{i=1}^n E_i \right)^2 - \left( \sum_{i=1}^n p_i \right)^2 \right\}^{1/2}
= m \left\{ n + 2 \sum_{1 \le i < j \le n} \cosh (\theta_i - \theta_j) \right\}^{1/2}
\ ,
\end{gather}
where trivial hyperbolic function identities are applied. Notably the invariant mass depends only on the rapidity differences, so that it is convenient to define the new variables
\begin{gather}
	u_{i} = \theta_i - \theta_{i+1}, \qquad {\rm for}\qquad  i =1,\dots, n-1\ .
\end{gather}
The kinematics of the $n$-particle system are completely specified by the set of variables $\theta_{i=1,\dots,n}$, or equivalently by $\theta_n$ and $u_{i=1,\dots,n-1}$. In terms of the new variables
\begin{gather}
\theta_i - \theta_j = \sum_{k=i}^{j-1} u_k \ , \qquad \text{for } 1 \le i < j \le n \ .
\label{eq:app:deltatheta}
\end{gather}
Once the phase-space integral appearing in the definition of the $n$-particle contribution to the spectral density has been written in terms of the new variables, one can use the $\delta$-function over the spatial momentum to eliminate the integral over $\theta_n$. One is left with the integral over the variables $u_{i=1,\dots,n-1}$ of a known function times the $\delta$-function over the invariant mass
\begin{gather}
\rho_\text{BN}^{(n)}(\mu) = \int_0^\infty \frac{du_1 \cdots du_{n-1}}{(4\pi)^{n-1}} F^{(n)}(u) \delta \left( \mu - M^{(n)}(u) \right)
\label{eq:app:integral-delta}
\ .
\end{gather}
The form factor is parametrized as 
\begin{gather}
F^{(n)}(u) = \frac{\pi^{3n-2}}{4} G^{(n)}(u) \prod_{1 \le i < j \le n} \left| \frac{\theta_{ij} - i \pi}{ \theta_{ij} ( 2\pi i - \theta_{ij} ) } \tanh^2 \frac{\theta_{ij}}{2} \right|_{\theta_{ij} = \theta_i - \theta_j}
\ ,
\end{gather}
where the r.h.s. can be written as a function of $u$ by means of eq.~\eqref{eq:app:deltatheta}, and the functions $G^{(n)}(u)$ are polynomials in $u$ which are given explicitly in Ref.~\cite{Balog:1996ey} for $n=2,4,6$. We report here only
\begin{gather}
G^{(2)}(u) = 2
\ , \\
G^{(4)}(u) = -4 [ 6 \tau_2^3 + 9 \tau_3^2 + 40 \tau_2 \tau_4 ] + 8 \pi^2 [ 25 \tau_2^2 + 44 \tau_4 ] - 448 \pi^2 \tau_2 + 272 \pi^6
\ ,
\end{gather}
where, for the $n=4$ case, the following auxiliary variables have been
introduced
\begin{gather}
\tau_{k} = \sum_{1 \le i_1 < \dots < i_k \le 4}
\prod_{\ell=1}^k \left( \theta_{i_\ell} - \frac{1}{4} \sum_{j=1}^4 \theta_j \right)
	\quad {\rm for} \quad k = 1,2,3,4.
\end{gather}
The corresponding expressions for $n=6$ are considerably longer, but present no logical complication. In order to calculate the integral~\eqref{eq:app:integral-delta} numerically one can solve the equation $M^{(n)}(u) = \mu > 0$ for e.g. $u_{n-1}$. The only positive solution $\bar{u}_{n-1}(u_1,\dots,u_{n-2})$ can be written explicitly by means of standard algebraic manipulations. The integral over $u_{n-1}$ is then explicitly given by
\begin{gather}
\rho_\text{BN}^{(n)}(\mu)
=
\int_{M^{(n)}(u_1,\dots,u_{n-2},0)<\mu} \frac{du_1 \cdots du_{n-2}}{(4\pi)^{n-1}} \left\{ \left[ \frac{\partial M^{(n)}}{\partial u_{n-1}}(u) \right]^{-1} F^{(n)}(u) \right\}_{u_{n-1} = \bar{u}_{n-1}(u_1,\dots,u_{n-2})}
\label{eq:app:integral}
\ .
\end{gather}
Notice that for $n=2$ there is no remaining integral, and one obtains the 
closed-form expression
\begin{gather}
\rho_\text{BN}^{(2)}(\mu)
=
\theta(\mu-2m)
\frac{8 \pi^3 m^5}{\mu^6}
\left( \frac{\mu^2}{4m^2} - 1 \right)^{5/2}
\left[
\frac{\bar{u}_1^2 + \pi^2}{\bar{u}_1^2 (4\pi^2 + \bar{u}_1^2)}
\right]_{\bar{u}_1 = 2 \cosh^{-1} \frac{\mu}{2m}}
\ .
\end{gather}
For $n=4,6$, the integral in~\eqref{eq:app:integral} is equivalently replaced by
\begin{gather}
\int_{M^{(n)}(u_1,\dots,u_{n-2},0)<\mu} \frac{du_1 \cdots du_{n-2}}{(4\pi)^{n-1}}
\ \longrightarrow \
\int_0^{\cosh^{-1} \frac{\mu^2 - n \, m}{2}}
\frac{du_1 \cdots du_{n-2}}{(4\pi)^{n-1}}
\,
\theta\left( \mu - M^{(n)}(u_1,\dots,u_{n-2},0) \right)
\ ,
\end{gather}
and is calculated numerically for $m=1$ (the $m$ dependence can be trivially reintroduced by dimensional analysis) and a selection of values of $\mu$ by means of the GNU Scientific Library implementation of Press and Farrar's MISER Monte Carlo algorithm~\cite{Press:1989vk}. In the region $\mu \le 40 m$, the $6$-particle contribution to the spectral density is always smaller than the $0.2\%$ of the sum of the $2$- and $4$-particle contributions. 

\section{Spectral reconstruction implementation}\label{a:hlt}
Explicit expressions and numerical implementation details of the spectral reconstruction algorithm are provided here for completeness. The presentation follows Ref.~\cite{Hansen:2019idp} with a slightly different notation. Additional information concerning the constraints used in this work is also provided.

By using a matrix-vector notation, the functionals in Eq.~\eqref{eq:AandB} can be expressed as
\begin{gather}
A[g] =
a\left\{\boldsymbol{g}^T \cdot\boldsymbol{A}\cdot \boldsymbol{g} -2\boldsymbol{g}^T\cdot \boldsymbol{f}\right\} + A[0]\;,
\qquad
B[g] =
\boldsymbol{g}^T \cdot \boldsymbol{B}\cdot \boldsymbol{g} \;,
\end{gather}
where $\boldsymbol{g}^T=(g_1,\cdots,g_{\tau_{\rm max}})$ is the vector collecting the coefficients $g_\tau$, $\boldsymbol{f}$ the vector with entries
\begin{gather}
f_\tau=\int_{E_0}^\infty d\omega\, {\delta}_{\epsilon}(E,\omega)\, b_\tau(\omega) \;,
\label{eq:fdef}
\end{gather}
and $\boldsymbol{A}$ and $\boldsymbol{B}$ matrices with elements
\begin{gather}
A_{\tau\tau^\prime}=a\int_{E_0}^\infty d\omega\, b_\tau(\omega)\, b_{\tau^\prime}(\omega)\;,
\qquad
B_{\tau\tau^\prime}= {\rm Cov} \left[aC(a\tau), aC(a\tau^\prime) \right]\;.
\end{gather}
Here $C(a\tau)$ is the correlator at time $\tau$ in lattice units and
\begin{gather}
A[0]=\int_{E_0}^\infty d\omega\, \left\{{\delta}_{\epsilon}(E,\omega)\right\}^2 \;.
\label{eq:A0def}
\end{gather}
With these definitions the vector $\boldsymbol{g^\lambda}$ that solves the minimum conditions of Eq.~\eqref{eq:Wminimumg} is given by
\begin{gather}
\boldsymbol{g^\lambda} = (1-\lambda)\boldsymbol{W^{-1}_\lambda}\cdot \boldsymbol{f}\;,
\qquad
\boldsymbol{W_\lambda}=(1-\lambda)\boldsymbol{A}+\lambda \frac{A[0]}{a}\boldsymbol{B}\;.
\label{eq:sol1}
\end{gather}

In Fig.~\ref{f:const} the minimization problem is subject to two different constraints on the reconstructed kernel. The first is detailed in Ref.~\cite{Hansen:2019idp} and forces the reconstructed and target kernels to have the same area
\begin{gather}
\int_{E_0}^\infty d\omega\, a\sum_{\tau =1}^{\tau_{\rm max}} g_\tau \, b_{\tau}(\omega)
=
\int_{E_0}^\infty d\omega\, {\delta}_{\epsilon}(E,\omega)\;.
\label{eq:cost1}
\end{gather}
The second constraint requires the reconstructed and target kernels to coincide at $\omega=E$
\begin{gather}
a\sum_{\tau =1}^{\tau_{\rm max}} g_\tau \, b_{\tau}(E)
=
{\delta}_{\epsilon}(E,E)\;.
\label{eq:cost2}
\end{gather}
Both constraints are represented in vector notation as
\begin{gather}
\boldsymbol{R}^T\cdot \boldsymbol{g} = r\;,
\label{eq:costgen}
\end{gather}
where for the constraint in Eq.~\eqref{eq:cost1} the entries of the vector $\boldsymbol{R}$ and the constant $r$ are given by
\begin{gather}
R_\tau=a\int_{E_0}^\infty d\omega\, b_\tau(\omega) \;,
\qquad
r=\int_{E_0}^\infty d\omega\, {\delta}_{\epsilon}(E,\omega)\;,
\label{eq:Rdef}
\end{gather}
while for the constraint in Eq.~\eqref{eq:cost2} they are given by
\begin{gather}
R_\tau=b_\tau(E) \;,
\qquad
r=\frac{{\delta}_{\epsilon}(E,E)}{a}\;.
\end{gather}
The solution of the minimization problem subject to Eq.~\eqref{eq:costgen} is
\begin{gather}
\boldsymbol{g^\lambda} = (1-\lambda)\boldsymbol{W^{-1}_\lambda}\cdot \boldsymbol{f}
+
\boldsymbol{W^{-1}_\lambda}\cdot \boldsymbol{R}\,
\frac{r-(1-\lambda)\boldsymbol{R}^T\cdot \boldsymbol{W^{-1}_\lambda}\cdot \boldsymbol{f}}{\boldsymbol{R}^T\cdot \boldsymbol{W^{-1}_\lambda}\cdot \boldsymbol{R}}\;.
\label{eq:sol2}
\end{gather}

For the basis functions actually used in this work, namely
\begin{gather}
b_{T,\tau}(E)= e^{-aE\tau} + e^{-E(T-a\tau)}\;,
\end{gather}
explicit expressions for the entries of the matrix $\boldsymbol{A}$ are
\begin{gather}
A_{\tau\tau^\prime} = \frac{e^{-E_0(a\tau + a\tau^\prime)}}{\tau + \tau^\prime} +\frac{e^{-E_0(T-a\tau + a\tau^\prime)}}{T/a - \tau + \tau^\prime}
+ \frac{e^{-E_0(T-a\tau^\prime + a\tau)}}{T/a-\tau^\prime + \tau} + \frac{e^{-E_0(2T - a\tau - a\tau^\prime)}}{2T/a - \tau - \tau^\prime}\;.
\end{gather}
When $\tau_{\rm max}$ is large this matrix is poorly conditioned. Consequently, when both $\lambda$ and the errors on the input data are small, the coefficients $g_\tau^\lambda$ are large in magnitude and oscillate in sign at different values of $\tau$. In these situations the signal on the reconstructed spectral density, given in vector notation by
\begin{gather}
\rho^\lambda_\epsilon(E)=\boldsymbol{c}^T\cdot \boldsymbol{g^\lambda}\;,
\qquad
c_\tau=a C(a\tau)\;,
\end{gather}
results from large numerical cancellations. For this reason (as in Ref.~\cite{Hansen:2019idp}), an implementation using extended-precision arithmetic is advocated. This is relatively straightforward for the kernels considered in this work using the libraries of Refs.~\cite{Johansson2017arb,MPFR} because the integrals appearing in Eqs.~\eqref{eq:fdef}, \eqref{eq:A0def}, and \eqref{eq:Rdef} can be expressed in terms of standard special functions already implemented in these packages\footnote{See for example the publicly available code at \url{https://github.com/mrlhansen/rmsd}}. In the case of more generic kernels a numerical integration implemented in extended-precision arithmetic may be required. An implementation based on the double-exponential quadrature algorithm~\cite{MORI2001287} is available upon request.

\section{Finite-volume effects in a model smeared spectral density}\label{s:app_fv}

In this appendix we present results concerning finite-volume effects in an idealized system where the spectral density is given entirely by $\rho^{(2)}(E)$ in 
Eq.~\eqref{e:2part}  and its finite-volume analog contains only two-particle states.\footnote{Stated more precisely, this condition means that the finite-volume smeared spectral density is equal to the one in Eq.~\eqref{eq:rho2FV}.} The results are therefore relevant for $\rho_{\epsilon}(E)$ in the O(3) model for values of $E$ and $\epsilon$ for which $n\ge 4$ particle states do not significantly contribute. Using the notation of Sec.~\ref{s:fv}, we first define $\rho_{\infty, L, \epsilon}(E)$ as the full spectral density (with $n=2,4,6, \cdots$ contributions) at finite $L$ and infinite $T$ smeared with a resolution function. This can be written as
\begin{equation}
\label{eq:fullSmearedFVSF}
\rho_{\infty, L, \epsilon}(E) = \sum_{n} c_n(L)\, \delta_{\epsilon}(E - E_n(L)) \,,
\end{equation}
where $\delta_{\epsilon}(x)$ a generic smearing kernel, and
\begin{equation}
c_n(L) = L \sum_a \big \vert \langle 0 \vert \hat j^a_1(0) \vert n \rangle_L \big \vert^2 \,.
\end{equation}

In absence of $n\ge 4$ particle states in the sum in Eq.~\eqref{eq:fullSmearedFVSF}  the $E_n(L)$ and $c_n(L)$ are given by the L{\"u}scher quantization condition~\cite{Luscher:1990ck} and the Lellouch-L{\"u}scher relation \cite{Lellouch:2000pv}, respectively\footnote{See also Ref.~\cite{Briceno:2020rar} for explicit expressions relevant to a two-dimensional spacetime.}
\begin{align}
\label{eq:energyQC}
E_n(L) & = Q^{-1}(\pi n) + O({\rm e}^{- m L}) \,, \\
c_n(L) & = \pi \bigg ( \frac{\partial Q(E) }{\partial E} \bigg )^{-1} \rho^{(2)}(E) \bigg \vert_{E = E_n(L)} + O({\rm e}^{- m L}) \,,
\end{align}
with
\begin{equation}
\label{e:qdef}
Q(E) \equiv \delta_{I=1}(E) + \frac{L \sqrt{E^2/4-m^2} }{2} \,.
\end{equation}

This means that (up to corrections exponentially suppressed in $L$)  both the energies and overlaps are dictated by kinematic factors and the two-particle scattering phase shift $\delta_{I=1}(E)$. The $I=1$ label emphasizes that this scattering phase shift corresponds to two-particle states transforming irreducibly under the fundamental representation of the global O(3) symmetry and distinguishes the phase shift from the resolution function $\delta_{\epsilon}(x)$ and the Dirac $\delta$-function. The phaseshift $\delta_{I=1}$ can be written explicitly via the analytically-known $S$-matrix, denoted $S(k)$
\begin{align}
\delta_{I=1}(E) & \equiv \frac{1}{2i} \log [ S(k) ]\bigg \vert_{k = \sqrt{E^2/4-m^2}}, \qquad
S(k)  = \frac{\theta + 2 i \pi}{\theta - 2 i \pi} \frac{\theta - i \pi}{\theta + i \pi} \bigg \vert_{\theta = 2 \sinh^{-1} \frac{k}{m}}\,,
\label{e:smatdef}
\end{align}
where the log is defined such that $0 < \delta_{I=1}(E) < \pi$. Note that the $\theta$ used here matches that of Eq.~\eqref{e:2part}, since
\begin{equation}
\frac{E^2}{4 m^2} = 1 + \frac{k^2}{m^2} \ \Longrightarrow \ \sinh^{-1} \frac{k}{m} = \cosh^{-1} \frac{E}{2m} \,.
\end{equation}

The two-particle, finite-volume, smeared spectral density is therefore defined as
\begin{equation}
\label{eq:rho2FV}
\rho^{(2)}_{L, \epsilon}(E) \equiv \sum_n \pi \bigg ( \frac{\partial Q(\omega) }{\partial \omega} \bigg )^{-1} \delta_{\epsilon}(E - \omega) \rho^{(2)}(\omega) \bigg \vert_{\omega = Q^{-1}(\pi n)} \,,
\end{equation}
where we have dropped the $T=\infty$ label. If $\delta_{\epsilon}(x)$ has compact support and $\epsilon$ and $E$ are chosen so that only $E_n(L)<4m$ states contribute, then this corresponds exactly to the smeared finite-volume spectral density in the full theory. By contrast, for non-compact smearing or for $E > 4m$, $\rho^{(2)}_{L, \epsilon}(E)$ is an approximation in which the role of infinite-volume states with more than two-particles is ignored.

\bigskip

In the remainder of this appendix, we show that terms scaling as $1/L$ cancel in  $\rho^{(2)}_{L, \epsilon}(E)$ provided that $\delta_{\epsilon}(x)$ is differentiable and falls off fast enough as $x \to \infty$. The precise condition is given after Eq.~\eqref{eq:invLcancels} below. We additionally show the stronger result that $\rho^{(2)}_{L, \epsilon}(E)$ has exponentially suppressed volume effects provided $\delta_{\epsilon}(x)$ is analytic in some finite-width strip about the real axis.
Before investigating the nature of the scaling at asymptotically large $L$, we
formally evaluate the $L \to \infty$ limit of Eq.~\eqref{eq:rho2FV}
\begin{align}
\rho^{(2)}_{\infty, \epsilon}(E) & \equiv \lim_{L \to \infty} \rho^{(2)}_{L, \epsilon}(E) \,, \\
& = \lim_{L \to \infty} \frac{1}{L} \sum_n \pi \bigg ( \frac{1}{2} \frac{\partial \sqrt{\omega^2/4-m^2} }{\partial \omega} + \frac{1}{L} \frac{\partial \delta_{I=1}(\omega) }{\partial \omega} \bigg )^{-1} \delta_{\epsilon}(E - \omega) \rho^{(2)}(\omega) \bigg \vert_{\omega = Q^{-1}(\pi n)} \,, \\
& = \int_0^\infty \frac{d k}{2 \pi} \, \pi \bigg ( \frac{1}{2} \frac{\partial \sqrt{\omega^2/4-m^2} }{\partial \omega} \bigg )^{-1} \delta_{\epsilon}(E - \omega) \rho^{(2)}(\omega) \bigg \vert_{\omega = 2 \sqrt{m^2 + k^2}} \,,
\end{align}
where in the final line we have dropped contributions to the finite-volume energy and the Lellouch-L{\"u}scher factor that are $1/L$ suppressed, replaced the sum over $n$ with an integral and performed the change of variables $k = 2 \pi n/L$. These results hold for any smearing kernel $\delta_{\epsilon}(x)$ for which the integral converges. Finally, changing integration variables to $\omega$ yields
\begin{align}
\rho^{(2)}_{\infty, \epsilon}(E) & = \int_{2m}^\infty d \omega \, \delta_{\epsilon}(E - \omega) \rho^{(2)}(\omega) \,,
\end{align}
as expected.

\subsection{$1/L$ cancellation}

We now show that $1/L$ contributions to the finite-volume energies and Lellouch-L{\"u}scher factors cancel in the definition of $\rho^{(2)}_{L, \epsilon}(E)$. First write
\begin{equation}
\rho^{(2)}_{L, \epsilon}(E) = \rho^{(2)}_{\infty, \epsilon}(E) + \frac{ c^{(1)}(E)}{L} + \mathcal O(1/L^2) \,,
\end{equation}
implying
\begin{equation}
c^{(1)}(E) \equiv \lim_{L \to \infty} L \big [\rho^{(2)}_{L, \epsilon}(E) - \rho^{(2)}_{\infty, \epsilon}(E) \big ] \,.
\end{equation}

Contributions to this coefficient arise from the leading-order shift to both the matrix element and the energy. The former can be written as
\begin{equation}
\bigg ( \frac{\partial Q(\omega) }{\partial \omega} \bigg )^{-1} = \frac{1}{L} \frac{8 k}{\omega} \bigg [ 1 - \frac{1}{L} \frac{8 k}{\omega} \frac{\partial \delta_{I=1}(\omega) }{\partial \omega} \bigg ] + \mathcal O(1/L^3) \,,
\end{equation}
where we have substituted $ \partial k / \partial \omega = \omega/(4 k)$ for $k = \sqrt{\omega^2/4-m^2} $.
The corresponding result for the energy is obtained by expanding Eq.~\eqref{eq:energyQC}
\begin{equation}
E_n(L) = 2 \sqrt{m^2 + k^2} - \frac{1}{L} \frac{8 k}{\omega} \delta_{I=1}(\omega) + \mathcal O(1/L^2) \,,
\end{equation}
where here $k = 2 \pi n/L$.

Putting everything together yields
\begin{equation}
c^{(1)}(E) = - \lim_{L \to \infty} \frac{1}{L} \sum_n \pi \frac{8 k}{\omega} \bigg ( \delta_{I=1}(\omega) \frac{\partial}{\partial \omega} + \frac{\partial \delta_{I=1}(\omega) }{\partial \omega} \bigg ) \frac{8 k}{\omega} \delta_{\epsilon}(E - \omega) \rho^{(2)}(\omega) \bigg \vert_{ \omega = 2 \sqrt{m^2 + k^2}} \,,
\end{equation}
where the first term in parenthesis arises from the energy shift and the second from the matrix element. Evaluating the limit gives
\begin{equation}
\label{eq:invLcancels}
c^{(1)}(E) = - \int_{2 m}^\infty d \omega \, \frac{d}{d \omega} \bigg [ \delta_{I=1}(\omega) \frac{8 \sqrt{\omega^2/4-m^2}}{\omega} \delta_{\epsilon}(E - \omega) \rho^{(2)}(\omega) \bigg ] \,.
\end{equation}
The precise condition on $\delta_{\epsilon}(x)$ is therefore that the integral above should vanish. Denoting the quantity in square brackets by $f(\omega)$, this holds if $d f(\omega)/d\omega$ is integrable, $f(\omega)$ is differentiable everywhere, and $f(\omega)$ vanishes at both $\omega = 2 m$ and infinity. All the smearing kernels considered in this work satisfy this condition as do much more aggressive choices, such as $ \delta_{\epsilon}(x) $ with compact support in the region $[x-\epsilon, x+\epsilon]$. 

\subsection{Exponentially suppressed volume effects}

We now demonstrate that for a more restrictive class of smearing kernels, finite-volume effects are indeed exponentially suppressed in $L$. This result also holds for all of the smearing kernels used in this work.
The argument is based on an elegant identity that is closely related to the derivation of the Lellouch-L{\"u}scher formalism. One can combine the appearance of $(\partial Q(\omega)/\partial \omega)$ with the definition of the finite-volume energies to re-write Eq.~\eqref{eq:rho2FV} as
\begin{equation}
\rho^{(2)}_{L, \epsilon}(E) \equiv 2 \pi \int_{2m}^{\infty} \! d \omega \ \delta_{\epsilon}(E - \omega) \ \sum_{n = 0}^\infty \delta \big ( 2 Q(\omega) - 2 \pi n \big ) \rho^{(2)}(\omega) \,,
\end{equation}
where the '$\delta$' in the sum over $n$ denotes the Dirac $\delta$-function.
The relation $\delta \big ( f(x) \big) = \vert f'(x) \vert^{-1} \delta(x)$ can be used to show that this expression is equivalent to Eq.~\eqref{eq:rho2FV}.

Next we use that $Q(E)$ is non-negative\footnote{This holds for all $L$ since the phase shift $\delta_{I=1}$ is positive. For systems with negative a scattering phase shift $Q(E)>0$ is only guaranteed above some sufficiently large value of $L$.} to extend the sum over $n$ to include all integers. Then applying the Poisson summation formula
\begin{equation}
\sum_{n = -\infty}^{\infty} \delta(x - 2 \pi n) = \frac{1}{2\pi} \sum_{n = - \infty}^\infty e^{i n x} \,
\end{equation}
gives
\begin{align}
\rho^{(2)}_{L, \epsilon}(E) & = \sum_{n = - \infty}^{\infty} \int_{2m}^{\infty} \! d \omega \, \delta_{\epsilon}(E - \omega) \, e^{i 2 n Q(\omega)} \rho^{(2)}(\omega) \,, \\
& = \sum_{n = - \infty}^{\infty} \int_{2m}^{\infty} \! d \omega \, \delta_{\epsilon}(E - \omega) \, e^{i n L k} [S(k)]^n \rho^{(2)}(\omega) \,,
\end{align}
where in the second line we have substituted the definition of $Q(E)$ from Eq.~\eqref{e:qdef}.

Taking the expressions for the spectral density and $S$-matrix (Eqs.~\eqref{e:2part} and \eqref{e:smatdef} respectively) and changing the variable of integration to $k=\sqrt{\omega^2/4-m^2}$ now results in 
\begin{align}
\label{e:rhoLpreshift}
&
\rho^{(2)}_{L, \epsilon}(E) = \frac{3\pi^3}{8} \sum_{n = - \infty}^{\infty} \int_{-\infty}^\infty d k \, \delta_{\epsilon}\Big(E - 2\sqrt{m^2+k^2}\Big) \, e^{ i n L k } \, \mathcal I_n(k) \,, 
\\
&
\mathcal I_n(k) \equiv \left. \bigg [ \frac{\theta + 2 i \pi}{\theta - 2 i \pi} \frac{\theta - i \pi}{\theta + i \pi} \bigg]^n \frac{1}{\theta^2}
\,
\frac{\theta^2 + \pi^2}{\theta^2 + 4\pi^2} \tanh^4 \frac{\theta}{2} \right|_{\theta = 2 \sinh^{-1} (k/m)}
\,.%
\label{eq:Fsum}
\end{align}
We have also used the invariance of the integrand under the simultaneous replacements $n \to - n$ and $k \to - k$ ($\theta \to - \theta$) to `unfold' the integral to the whole real axis. The function $\mathcal I_n(k)$ is analytic in the whole complex plane except for two cuts on $\pm i [m,\infty)$.

Assuming that $\delta_{\epsilon}\left(E - 2\sqrt{m^2+k^2}\right)$ is analytic in the complex strip $\text{Im}\, k < \mu$ for some $\mu \le m$, one can shift the integration contour to $\mathbb R + i \mu$. Concerning the kernels used in this work, $\mu = m$ always holds for the Gaussian kernel ${\sf g}$ but for the Cauchy kernels, ${\sf c0}$, ${\sf c1}$ and ${\sf c2}$, the singularity at $(E - 2 \sqrt{m^2 + k^2})^2 = - \epsilon^2$ can be a distance less than $m$ from the real axis. In this case one has
\begin{equation}
\label{e:mudef}
\mu = \text{Im} \sqrt{\frac{E^2 - 4m^2 - \epsilon^2}{4} + \frac{ i E \epsilon}{2}}
= \frac{ E \epsilon}{2\sqrt{E^2 - 4m^2}} + O(\epsilon^2)
\,,
\end{equation}
where the small $\epsilon$ expansion is included for illustration. The following formulae are also only valid if the segments at infinity do not contribute. We have confirmed this is the case for the kernels used in this work.

Extending the integration range and shifting the contour in Eq.~\eqref{e:rhoLpreshift} as described, we find that the difference between finite- and infinite-volume smeared spectral densities can be written as
\begin{equation}
\label{e:aftershift}
\Delta \rho^{(2)}_{L, \epsilon}(E) \equiv \rho^{(2)}_{L, \epsilon}(E) - \rho^{(2)}_{\infty, \epsilon}(E) = \sum_{n = 1}^{\infty} \mathcal C_n(E, L, \mu) e^{ - n \mu L } \,.
\end{equation}
The coefficient $ \mathcal C_n$ is oscillatory with an envelope that is either constant or falling with a power of $L$. It is defined as
\begin{equation}
\mathcal C_n(E, L, \mu) \equiv \frac{3\pi^3}{8} \lim_{\eta \to \mu^-} \text{Re} \int_{-\infty}^\infty d x \, \delta_{\epsilon}(E - \omega_{\eta}(x)) \, e^{ i n L x } \, \mathcal I_n(\theta_{\eta}(x)) \,,
\end{equation}
with $ \theta_{\eta}(x) = 2 \sinh^{-1} [(x + i \eta)/m]$ and $\omega_{\eta}(x) =2 \sqrt{(x+i\eta)^2 + m^2}$.

%%%%%%%%%%%%%%%%
%%           FIGURE 10          %%
%%%%%%%%%%%%%%%%
\begin{figure}
\includegraphics[width=\textwidth]{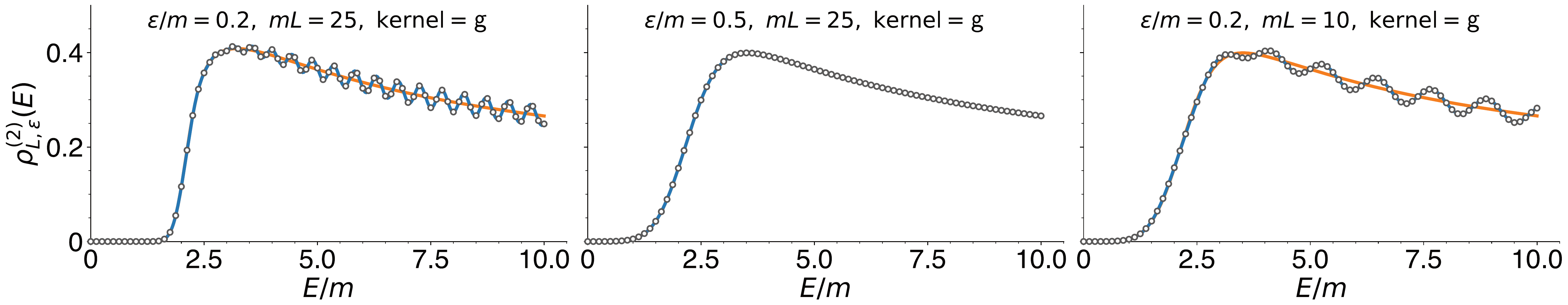} 
\caption{\label{f:fveffectsvsE}Theoretically predicted two-particle spectral densities for three sets of $\epsilon/m$ and $mL$ values (as indicated) with the Gaussian smearing kernel. Each panel includes the smeared infinite-volume results (orange), as well as the finite-volume smeared spectral density defined in Eq.~\eqref{eq:rho2FV} (blue). The points arise from numerically evaluating the $n=1$ term of Eq.~\eqref{e:aftershift} and combining with $\rho_{\infty, \epsilon}(E)$.}
\end{figure}
%%%%%%%%%%%%%%%%
%%%%%%%%%%%%%%%%
%%%%%%%%%%%%%%%%

%%%%%%%%%%%%%%%%
%%           FIGURE 11          %%
%%%%%%%%%%%%%%%%
\begin{figure}
\includegraphics[width=\textwidth]{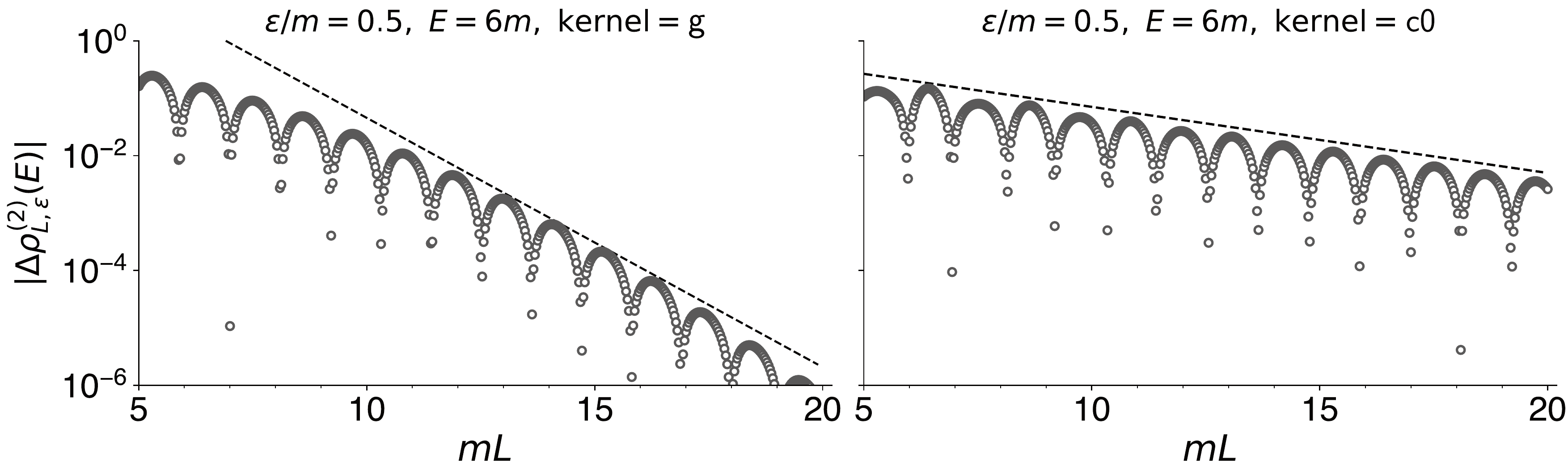}
\caption{\label{f:fveffectsvsL}Plots of the finite-volume residue vs.~$L$ for the Gaussian kernel (${\sf g}$, left) and Cauchy kernel (${\sf c0}$, right), determined using  Eq.~\eqref{e:aftershift}. The dashed lines give the predicted $e^{- \mu L}$ scaling for each (with a pre-factor chosen to position the curve for ease of comparison); $\mu = m$ for the Gaussian and $\mu = 0.265 m$ for the Cauchy, slightly exceeding $\epsilon/2$ as given by Eq.~\eqref{e:mudef}. The Gaussian data includes only the $n=1$ term of Eq.~\eqref{e:aftershift} while the Cauchy includes the sum over the $n=1,2,3$ terms, i.e.~all terms that fall-off slower than $e^{- mL}$.}
\end{figure}
%%%%%%%%%%%%%%%%
%%%%%%%%%%%%%%%%
%%%%%%%%%%%%%%%%

We have also studied these results numerically and confirmed that the straightforward expression based on a sum over finite-volume states in Eq.~\eqref{eq:rho2FV} matches the sum over Poisson modes evaluated both with the original contour (Eq.~\eqref{e:rhoLpreshift}) and the shifted contour which makes the $e^{- \mu L}$ scaling manifest (Eq.~\eqref{e:aftershift}). In Fig.~\ref{f:fveffectsvsE} we show the results for the Gaussian kernel plotted as a function of $E$ for various values of fixed $\epsilon$ and $L$. In Fig.~\ref{f:fveffectsvsL} we plot $\vert \Delta \rho^{(2)}_{L, \epsilon}(E)\vert $ at fixed $E$ and $\epsilon$ versus $L$ for both the Gaussian and Cauchy kernels. We see that the volume effects are oscillatory functions of $L$ with an envelope decaying according to the predicted exponential. 
The derivations of this appendix therefore suggest that finite-volume effects are small for the numerical results in Sec.~\ref{s:fixed}.  

\section{Simulation algorithm}\label{a:alg}

We employ the single-cluster algorithm and associated cluster estimators from Ref.~\cite{Luscher:1990ck} which are briefly summarized here. A cluster update
proceeds as follows. First,
a random vector $r \in {\mathbb R}^3, \, |r| = 1$ is drawn uniformly from the unit sphere. Then a `seed' site is chosen uniformly as the first member of the cluster. For each new site $x$ added to the cluster, consider all non-cluster sites among the four nearest neighbors of $x$. A neighbor $y$ is added to the cluster with probability
\begin{align}
 p_{\rm add} = 1 - \exp\left[ \min\{-2\beta \sigma_{r}(x) \sigma_r(y), 0\}\right]
\end{align}
where $\sigma_r(x) = \sigma(x) \cdot r$. After all cluster neighbors have been considered for addition, all cluster sites are updated according to
\begin{align}
\sigma^{a}(x) \rightarrow \sigma^a(x) - 2\sigma_r(x) r^a.
\end{align}
In order to employ a cluster estimator for $C(a\tau)$, which contains four fundamential fields, a second orthogonal cluster update is required. This proceeds by
choosing a second random vector $u$ from the unit sphere with the constraint that $r \cdot u = 0$ and then performing a second cluster update in the same 
manner.
The combination of a single $r$-update followed by a $u$-update is henceforth
referred to as a `cluster update', the number of which are tabulated in Tab.~\ref{t:ens} for each ensemble.

The cluster estimator for $C(a\tau)$ is built from
\begin{gather}
\tilde{\sigma}_{r,u}(x) = \sqrt{\frac{|\Lambda|}{N_{r,u}}}\theta_{r,u}(x) \sigma_{r,u}(x)
\end{gather}
where $N_{r,u}/|\Lambda|$ is the cluster fraction and $\theta_{r,u}(x)$ the characteristic function. Expectation values $\langle \dots \rangle_{\rm 1C}$ denote integration over all possible pairs of orthogonal single-cluster configurations as well as the usual integration over the field variables.
Straightforward application
of Eq.~\eqref{e:lat_cur} gives
\begin{align}\label{e:j_man}
\langle j^{a}_{\mu}(x) j^{a}_{\mu}(y)\rangle =
2 \beta^2 \, P(ab|cd) \, \langle \sigma_a(x)\sigma_b(x+a\hat{\mu})\sigma_c(y)\sigma_d(y+a\hat{\mu}) \rangle = 12\beta^2\, \langle \tilde{\sigma}_{[r}(x)\tilde{\sigma}_{u]}(x+a\hat{\mu})\,
\tilde{\sigma}_{[r}(y)\sigma_{u]}(y+a\hat{\mu})\rangle_{\rm 1C}
\end{align}
where $P(ab|cd) = \frac{1}{2}(\delta_{ac}\delta_{bd} - \delta_{ad}\delta_{bc})$
and $\tilde{\sigma}_{[r}(x)\tilde{\sigma}_{u]}(y) = \frac{1}{2}\{\tilde{\sigma}_{r}(x)\tilde{\sigma}_u(y) - \tilde{\sigma}_u(x)\tilde{\sigma}_r(y)\}$. The correlator is then given by
\begin{align}\label{e:cc}
C(a\tau) = \frac{12\beta^2}{L/a} \langle \Phi_{ru}(a\tau) \, \Phi_{ru}(0) \rangle_{\rm 1C},
\qquad \Phi_{ru}(a\tau) = \sum_{\boldsymbol{x}} \tilde{\sigma}_{[r}(a\tau, \boldsymbol{x})\tilde{\sigma}_{u]}(a\tau, \boldsymbol{x} + a\hat{1})\,,
\end{align}
where (although not denoted explicitly) time translation invariance is employed to average over all equivalent time separations on the periodic torus.

The estimator for $C(a\tau)$ in Eq.~\eqref{e:cc} is not positive definite, in contrast to the
cluster estimator for the two-point function of the fundamental fields, which requires only a single cluster.
The computational cost of $\Phi_{ru}(a\tau)$ for a single update scales only
weakly with the lattice size at fixed $\beta$, since only pairs of sites which are neighbors in the spatial direction and belong to different clusters contribute.
However this overlap
becomes increasingly unlikely, resulting in an increasing statistical error for a fixed number of cluster updates. Despite this `cluster cutoff', the variance
of the estimator in Eq.~\eqref{e:cc} decays exponentially with increasing $\tau$, while the one
for standard estimator (given by the middle expression in Eq.~\eqref{e:j_man}) approaches a constant. The cluster estimator thus results in a significant improvement in the signal-to-noise ratio, which empirically decays with a rate roughly similar to $m$.

\bibliography{latticen}
\end{document}